\documentclass[fleqn,usenatbib]{mnras}

\usepackage{newtxtext,newtxmath}

\usepackage[T1]{fontenc}
\DeclareRobustCommand{\VAN}[3]{#2}
\let\VANthebibliography\thebibliography
\def\thebibliography{\DeclareRobustCommand{\VAN}[3]{##3}\VANthebibliography}

\usepackage{graphicx}	
\usepackage{amsmath}	
\usepackage{physics}
\usepackage{hyperref}
\usepackage{multirow, tabularx}
\usepackage{array}
\usepackage[table, svgnames, dvipsnames]{xcolor}
\usepackage{makecell, cellspace, caption}
\usepackage{mathtools}
\usepackage{enumitem}
\usepackage{bm}         
\usepackage{url}
\usepackage{braket}
\usepackage{threeparttable}
\usepackage{orcidlink}
\usepackage[normalem]{ulem}

\newcommand\Kelvin{\,{\rm K}}
\newcommand\Mpc{\,{\rm Mpc}}

\newcommand\pc{\,{\rm pc}}
\newcommand\mpc{\, h^{-1}{\rm {Mpc}}}

\newcommand\Kpc{\, {\rm {kpc}}}

\newcommand\Myr{\,{\rm Myr}}
\newcommand\Gyr{\,{\rm Gyr}}

\newcommand\kms{\,{\rm {km\, s^{-1}}}}
\newcommand\Msun{{\rm M_\odot}}
\newcommand\msun{\, h^{-1}{\rm M_\odot}}

\newcommand\perccm{\, {\rm cm}^{-3}}

\newcommand\figem{\bf}                  
\newcommand\objname{\tt}
\newcommand\software{\sc}
\newcommand\ssc{\rm\scriptscriptstyle}  

\defcitealias{moTwophaseModelGalaxy2024}{MCW24}

\title[Two-phase formation of galaxies]{Two-phase formation of galaxies: 
the coevolution between galaxies and dark matter halos}

\author[Qinglin Ma et al.]{
    Qinglin Ma\orcidlink{0009-0000-5148-9457}, $^{1}$\thanks{E-mail: maql21@mails.tsinghua.edu.cn}
    Yangyao Chen\orcidlink{0000-0002-4597-5798} $^{2,3,4,5}$\thanks{E-mail: yangyaochen.astro@foxmail.com}
    and
    Houjun Mo\orcidlink{0000-0001-5356-2419}$^{6}$
\\
$^{1}$Department of Astronomy, Tsinghua University, Beijing 100084, China\\
$^{2}$School of Astronomy and Space Science, Nanjing University, Nanjing, Jiangsu 210093, China\\
$^{3}$Key Laboratory of Modern Astronomy and Astrophysics, Nanjing University, Ministry of Education, Nanjing, Jiangsu 210093, China\\ 
$^{4}$Department of Astronomy, University of Science and Technology of China, Hefei, Anhui 230026, China\\
$^{5}$School of Astronomy and Space Science, University of Science and Technology of China, Hefei, Anhui 230026, China\\
$^{6}$Department of Astronomy, University of Massachusetts, Amherst, MA 01003, USA
}

\date{Accepted XXX. Received YYY; in original form ZZZ}

\pubyear{\the\year{}}

\begin{document}
\label{firstpage}
\pagerange{\pageref{firstpage}--\pageref{lastpage}}
\maketitle

\begin{abstract}
We use FIRE-2 cosmological zoom-in hydrodynamic simulations to investigate the
co-evolution between Milky Way-size galaxies and their host dark matter halos.
We find that the formation of these galaxies follows a two-phase pattern,
with an early phase featured by hot dynamics, bulge-dominated structure 
and bursty star formation, and a later phase featured by cold dynamics,
disk-dominated structure and steady star formation. The transition times 
of these galaxy properties are correlated with the time
when the host halo transits from fast to slow accretion,
indicating the two-phase assembly of halos 
as a potential mechanism that drives the two-phase formation of galaxies.
The physical origin of dynamical hotness can be summarized into two modes 
of star formation: a scattered mode in which stars form 
at large radii within cold gas streams associated with fast assembly of halos,
and a concentrated mode in which stars form at small radii
through violent fragmentation from globally self-gravitated gas 
when halo assembly is about to slow down. 
Cold gaseous and stellar disks can form 
when the conditions of the two modes are removed by the stall of 
fast halo assembly and the reduction of gas by feedback processes.
The two modes of star formation leave distinct imprints on the structural
properties of high-redshift galaxies, providing implications to be tested by JWST 
and future observations.

\end{abstract}

\begin{keywords}
galaxies: kinematics and dynamics -- 
galaxies: formation -- 
galaxies: haloes -- 
galaxies: high-redshift
\end{keywords}



\section{Introduction} 
\label{sec:introduction} 

Galaxies exhibit diverse morphologies ranging from 
those dominated by bulges to those dominated by disks -- known as the Hubble 
sequence \citep[e.g.][]{Hubbleextragalactic1926,Hubblerene1936,SandageHubbleatlas1961,Kormendymorphological1979,Conseliceevolutiongalaxystructure2014}. 
Observations with resolved kinematic measurements suggest that the 
morphologies of galaxies are closely related to the motion of stars,
with bulges being populated mainly by stars with random orbits 
and disks with ordered rotation \citep[e.g.][]{Cappellaristructure2016,zhuOrbitalDecompositionCALIFA2018}.
Historically, these two types of motion were inferred to reflect the evolution
stages of galaxies, with ordered rotation being the early products of galaxy
formation owing to the collapse of gas with initial angular momentum in halos,
and random motion being generated by later `heating' processes such as mergers 
\citep[e.g.][]{hernquistStructureMergerRemnants1992,shenSizeDistributionGalaxies2003,
laceyUnifiedMultiwavelengthModel2016}. 
The formation of stable disks establishes the blue, star-forming sequence. Subsequent dynamical heating 
and quenching of star formation, driven by merger-induced bulge growth and supermassive black hole (SMBH) feedback, 
cause galaxies to migrate toward the red sequence, producing the observed bimodality in the galaxy population.
Such a picture of galaxy formation and 
morphological transformation has been widely adopted in theoretical models 
and achieved great success in reproducing the observed properties
of the galaxy population observed in the local Universe, such as the size-mass 
relations \citep[e.g.][]{shenSizeDistributionGalaxies2003,vandokkumConfirmationRemarkableCompactness2008,
vandokkumFormingCompactMassive2015,bezansonRelationCompactQuiescent2009,
newmanCanMinorMerging2012,grahamEllipticalDiskGalaxy2013,
vanderwel3DHST+CANDELSEvolutionGalaxy2014,somervilleRelationshipGalaxyDark2018,
millerNewViewSize2019,jiaSizeGrowthShort2024,songTransitionOutsideinInsideOut2025} and the bimodality
in the color-stellar mass relation \citep{stratevaColorSeparationGalaxy2001,baldryQuantifyingBimodalColorMagnitude2004}.

Within the $\Lambda$ cold dark matter ($\Lambda$CDM) paradigm, the formation 
and evolution of galaxies can be understood by starting from a set of initial
conditions set by the cosmological model and following the evolution 
by solving the gravitational and hydrodynamical coupling among different 
mass elements. One critical component in this framework is the formation of
the dark matter halos, which not only increases the density of baryonic matter 
so that subsequent cooling and condensation can produce the first generation of 
stars \citep[see e.g. the cooling diagram in fig. 2 of][]{moTwophaseModelGalaxy2024}, 
but also provides continuous fuel of gas and deep gravitational 
potential wells so that star formation can be sustained without being interrupted by
feedback processes \citep[e.g.][]{taylorEmergenceGlobularClusters2025,chenTwophaseModelGalaxy2025a}.
The assembly of dark matter halos thus provides the cosmological context
for understanding processes that eventually shape the diverse morphologies of 
galaxies observed.

Along this line, \citet[hereafter \citetalias{moTwophaseModelGalaxy2024}]{moTwophaseModelGalaxy2024}
proposed a `two-phase' scenario of galaxy formation, in which the two 
morphological components of galaxies, bulge and disk, naturally arise 
from cosmological structure formation. This scenario is motivated by the fact that the 
assembly of individual CDM halos follows a general pattern consisting of two phases: an early phase
of fast accretion and a later phase of slow accretion -- a discovery 
made decades ago in N-body simulations 
(\citealt{zhaoGrowthStructureDark2003}; see also \citealt{wechslerConcentrationsDarkHalos2002,zhaoAccurateUniversalModels2009,correaAccretionHistoryDark2015a}).
During the fast phase of a halo, rapid change of gravitational 
potential associated with the fast accretion of the halo, complemented by 
energy inputs from the gas accretion itself and feedback processes, can generate 
large perturbations in the gas
\citep{RobertsonAdiabaticHeatingContracting2012,MurrayCollapseSelfgravitatingTurbulent2017,
FieldingImpactStarFormation2017,GoldnerAccretionDrivenTurbulence2025,sunTurbulentFrameworkStar2026}, 
which can seed clouds of over-density and promote the formation of dense sub-clouds.
The rich amount of gas and the strong turbulence brought in
by the fast accretion imply that the fragmentation starts once the gas becomes 
self-gravitating and that the formation of dense sub-clouds proceeds rapidly before the 
gas can dissipate to form a dynamically cold disk. 
A galaxy formed in this phase is thus manifested as a self-gravitating gas 
cloud (SGC) filled with sub-clouds with random orbits, with star clusters 
born out of the sub-clouds inheriting their hot dynamics.
Dynamically cold gaseous and stellar disk can form after the host halo
transits to the slow phase and the gas content is reduced by feedback processes.
Implemented semi-analytically into subhalo merger trees,
this scenario can reproduce a wide range of observables,
and coherently predict the co-evolution among halos, galaxies, supermassive 
black holes (SMBHs), and globular clusters (GCs) from the seeding era to the present day
\citep[see][for details]{moTwophaseModelGalaxy2024,chenTwophaseModelGalaxy2024, 
chenTwophaseModelGalaxy2025,chenTwophaseModelGalaxy2025a}.

One important prediction of the two-phase scenario is an order in the build-up of the Hubble sequence: the dynamically hot (bulge) component of a 
galaxy is actually built first, and followed by the dynamically cold (disk) 
component, with a transition corresponding to that in halo assembly from fast to slow.
This ``upside-down'' order seems to be in conflict with the historical picture of galaxy formation and 
morphological transformation mentioned above, but becomes increasingly supported by 
hydrodynamic simulations \citep{dekelColdStreamsEarly2009,BrookFeedbackangular2012,
BirdInsideTracingAssembly2013,hopkinsFIRE2SimulationsPhysics2018,pillepichFirstResultsTNG502019,
ParkNewHorizonOrigin2019,yuBornThisWay2023,
Jung2025itagorahighresolutiongalaxy} 
and observations \citep{WuytsCANDELSSmoothstellar2012,KassinEpochdisksettling2012,SwinbankHiZELS2012,
WisnioskiKMOS3Dsurvey2015,SimonsEpochdiskassembly2017,PriceMOSDEFkinematic2020}. 
However, many questions remain unanswered.
For example, it is unclear whether the fast accretion can indeed sustain a gas-rich 
and turbulent SGC, whether sub-clouds can actually form in the SGC quickly before
settling into a rotation-supported disk, and whether feedback processes are strong enough to 
reduce the gas content to allow the formation of a stable disk. 
Given the challenges in obtaining a complete sample of high-redshift galaxies 
to perform dynamical decomposition, it is still difficult to answer these questions 
directly using observational data. On the other hand, modern zoom-in hydro simulations 
with cosmological initial conditions are able to resolve key processes involved in the
formation of individual galaxies, thus providing opportunities to test the 
viability of the two-phase model. 

In this paper, we use a set of such simulations performed as a part of the Feedback
In Realistic Environments (FIRE) project 
\citep{hopkinsFIRE2SimulationsPhysics2018} to analyze the co-evolution between galaxies 
and dark matter halos, focusing on the aforementioned questions. This paper is organized as follows. 
In \S\ref{sec:sim}, we describe the simulation and the method to quantify halos and galaxies.
In \S\ref{sec:co-evolution}, 
we show the two-phase growth pattern of galaxies, and demonstrate that 
it is consistent with being influenced by the two-phase assembly of dark matter 
halos.
In \S\ref{sec:driving-processes}, we analyze in detail the physical processes
that drive the two-phase growth of galaxies. In \S\ref{sec:early-growth}, we provide implications
for observations in a search for evidence induced by these processes.
Finally, we summarize and discuss our findings in \S\ref{sec:summaryanddiscussion}.

\section{The Simulation}
\label{sec:sim}

{
\renewcommand{\arraystretch}{1.2}
\begin{table*}
    \centering    
    \caption{{\figem The sample of FIRE-2 galaxies used in this paper}. 
    The properties listed in the columns are: 
    (i) the name of the galaxy, with the reference first introducing the galaxy 
    indicated in the parentheses (A, \citealt{GarrisonDwarfsatellitesabundence2019}; 
    B, \citealt{GarrisonDwarfsatellitessfr2019}; 
    C, \citealt{Garrisonsubhalo2017}; 
    D, \citealt{Wetzeldwarfgalaxy2016}; 
    E, \citealt{hopkinsFIRE2SimulationsPhysics2018}; 
    F, \citealt{SamuelFIREsatellite2020});
    (ii) $M_{\mathrm{vir}, \mathrm{dm}}$, the mass of dark matter enclosed in the 
    virial radius of the host halo at $z = 0$;
    (iii) $M_{*}$, stellar mass of the galaxy;
    (iv)  $R_{50, *}$, half-stellar-mass radius using all star particles linked to the galaxy;
    (v) $R_{\rm 50,bulge}$, half-stellar-mass radius using only bulge stars;
    (vi) $R_{90, *}$, the radius enclosing 90\% of the stellar mass, using 
    all star particles linked to the galaxy;
    (vii) $t_{\mathcal{B}}$, the transition time of burstiness;
    (viii) $t^{(\mathrm{start})}_{\mathcal{H}}$--$t^{(\mathrm{end})}_{\mathcal{H}}$, 
    the transitional interval of hotness;
    (ix) $t^{(\mathrm{start})}_{\mathrm{thin}}$--$t^{(\mathrm{end})}_{\mathrm{thin}}$,
    the transitional interval of thin-disk fraction;
    (x) $t_{\rm bulge,1/2}$, the half-mass formation time of stellar bulge;
    (xi) $t^{(\mathrm{start})}_{\gamma}$--$t^{(\mathrm{end})}_{\gamma}$,
    the transitional interval of halo accretion rate;
    (xii) $t_{\rm h, 1/2}$, the half-mass formation time of the host halo;
    (xiii) $t_{\text{h, Vpeak}}$, the epoch when the maximum circular velocity 
    of the halo reaches the peak during the assembly history.
    See \S\ref{sec:sim} for the details of the sample and the definitions 
    of the properties.
    }
    \label{tab:sample}
    \tabcolsep=0.16cm
    \begin{tabular}{l*{12}{c}}
        \hline
        \multirow{2}{*}{\bf Name (ref)}
        & $M_{\mathrm{vir}, \mathrm{dm}}$ 
        & $M_{*}$ 
        & $R_{\rm 50,*}$ 
        & $R_{\rm 50,bulge}$
        & $R_{\rm 90,*}$ 
        & $t_{\mathcal{B}}$ 
        & $t^{(\mathrm{start})}_{\mathcal{H}}$--$t^{(\mathrm{end})}_{\mathcal{H}}$ 
        & $t^{(\mathrm{start})}_{\mathrm{thin}}$--$t^{(\mathrm{end})}_{\mathrm{thin}}$ 
        & $t_{\rm bulge,1/2}$ 
        & $t^{(\mathrm{start})}_{\gamma}$--$t^{(\mathrm{end})}_{\gamma}$ 
        & $t_{\rm h, 1/2}$ 
        & $t_{\text{h, Vpeak}}$ 
    \\
        & [${\rm M}_{\odot}$] & [${\rm M}_{\odot}$] & [$\rm kpc$] & [$\rm kpc$] & [$\rm kpc$] 
            & [${\rm Gyr}$] & [${\rm Gyr}$] & [${\rm Gyr}$] & [${\rm Gyr}$] & [${\rm Gyr}$] & [${\rm Gyr}$] & [${\rm Gyr}$]    \\
        \hline
        \texttt{m12i} (D) & 6.3$\times$10$^{11}$ & 6.3$\times$10$^{10}$ & 2.8 & 1.6 & 9.1 
            & 10.6 & 7.2--10.4 & 7.0--10.4 & 5.6 & 4.5--7.0 & 4.7 & 5.7   \\
        
        \texttt{m12f} (C) & 8.9$\times$10$^{11}$ & 8.5$\times$10$^{10}$ & 3.8 & 1.6 & 14.3 
            & 8.9  & 5.9--8.7  & 5.7--8.7  & 4.5 & 2.5--6.8 & 6.0 & 3.5   \\
        
        \texttt{m12m} (E) & 8.1$\times$10$^{11}$ & 1.2$\times$10$^{11}$ & 5.0 & 1.7 & 12.6 
            & 10.1 & 8.1--11.4 & 2.7--11.4 & 7.0 & 5.2--7.9 & 5.5 & 3.4 \\
        
        \texttt{m12b} (A) & 7.6$\times$10$^{11}$ & 8.2$\times$10$^{10}$ & 2.8 & 1.4 & 10.6 
            & 7.6  & 6.1--6.9  & 6.3--6.9  & 5.1 & 3.0--6.8 & 5.0 & 8.4  \\
        
        \texttt{m12c} (A) & 7.5$\times$10$^{11}$ & 6.1$\times$10$^{10}$ & 3.4 & 1.7 & 9.8 
            & 10.3 & 6.6--11.3 & 6.8--11.3 & 5.5 & 3.7--9.4 & 7.9 & 4.9   \\
        
        \hline
        \texttt{m12w} (F) & 5.6$\times$10$^{11}$ & 5.5$\times$10$^{10}$ & 3.1 & 1.4 & 8.6 
            & 11.6 & 3.5--12.5 & 1.6--12.5 & 7.8 & 3.2--7.5 & 6.4 & 8.0  \\
        
        \texttt{m12r} (F) & 6.1$\times$10$^{11}$ & 1.8$\times$10$^{10}$ & 4.4 & 1.8 & 13.2 
            & 8.5  & 6.3--10.3 & 6.5--10.3 & 4.9 & 1.5--2.4 & 10.1 & 12.6  \\
        
        \texttt{m12z} (A) & 4.9$\times$10$^{11}$ & 2.1$\times$10$^{10}$ & 4.8 & 3.7 & 11.3 
            & -    & 6.4-- & 10.1-- & 9.5 & 3.5--11.8& 6.6 & 10.4  \\
        
        \hline
        \texttt{Romeo} (A) & 7.0$\times$10$^{11}$ & 7.3$\times$10$^{10}$ & 4.2 & 1.4 & 13.8 
            & 7.5  & 2.2--8.0  & 1.1--7.9  & 3.5  & 1.7--3.2 & 4.4 & 1.7  \\
        
        \texttt{Juliet} (A) & 5.8$\times$10$^{11}$ & 3.7$\times$10$^{10}$ & 3.7 & 1.5 & 13.8 
            & 9.3  & 5.9--10.3 & 5.9--10.3 & 4.4 & 1.8--3.5 & 6.3 & 2.5  \\

        \texttt{Thelma} (A) & 7.7$\times$10$^{11}$ & 7.7$\times$10$^{10}$ & 4.4 & 2.7 & 12.3 
            & 8.2  & 6.8--9.6  & 5.9--9.5  & 4.7 & 2.3--4.1 & 4.1 & 3.9 \\
        
        \texttt{Louise} (A) & 6.0$\times$10$^{11}$ & 2.7$\times$10$^{10}$ & 3.3 & 2.1 & 11.1 
            & 11.3 & 8.8--12.5 & 6.9--12.5 & 7.2 & 4.9--8.2 & 4.5 & 5.4  \\
        
        \texttt{Romulus} (B) & 1.1$\times$10$^{12}$ & 1.0$\times$10$^{11}$ & 4.6 & 1.5 & 17.4 
            & 9.5  & 5.2--8.8  & 4.9--8.8  & 5.0 & 4.1--7.0 & 4.3 & 5.1  \\
        
        \texttt{Remus} (B) & 6.9$\times$10$^{11}$ & 4.9$\times$10$^{10}$ & 3.3 & 1.7 & 11.6 
            & 8.1  & 4.5--9.5  & 4.6--8.8 & 3.9  & 2.2--4.4 & 4.6 & 4.9  \\
        
        \hline
    \end{tabular}
\end{table*}
}

In this paper, we use the cosmological hydrodynamic zoom-in simulations from FIRE-2,
a part of the FIRE project, performed using a multimethod gravity plus 
(magneto)hydrodynamics code GIZMO \citep{HopkinsGIZMO2015}, with updated
numerical methods and physical models as detailed in \citet{hopkinsFIRE2SimulationsPhysics2018}.
FIRE-2 adopted the mesh-free Lagrangian Godunov (MFM) method for hydrodynamics,
which provides adaptive spatial resolution while maintaining exact conservation of 
mass, energy, and momentum, excellent angular momentum conservation, 
and accurate shock capturing. 
Radiative heating and cooling for gas across a temperature range of 
$\sim 10 $--$10^{10}\Kelvin$ are included, involving 
free-free, photoionization/recombination, Compton, photo-electric, metal-line, 
molecular, fine-structure, dust collisional, and cosmic-ray processes.
Eleven elements (H, He, C, N, O, Ne, Mg, Si, S, Ca, Fe) are self-consistently 
tracked. Unresolved turbulent diffusion of metals in the gas is explicitly modeled 
\citep{Hopkinsturbulence2016,Sumetaldiffusion2017,Escalametaldiffusion2018}.
The simulations also include photoionization and heating from a redshift-dependent, 
spatially uniform ultraviolet (UV) background \citep{Faucherionizingbackground2009}
that reionizes the universe at $z \sim 10$. 
Collisionless star particles, representing ensembles of stars, are bred
from molecular gas satisfying self-gravitating, self-shielding, 
Jeans unstable conditions and exceeding a minimum density threshold 
of $1000 \perccm$ \citep{Krumholzmolecular2011}. 
FIRE-2 implemented distinct channels of stellar feedback, including that
from OB stars, AGB mass-loss, type Ia and type II supernovae, photoelectric heating, 
and radiation pressure \citep{Hopkinssupernova2018,Hopkinsrediativefeedback2020}. 
Such feedback regulates the star formation efficiency of star-forming gas
to a level of $\sim 1\%$--$10\%$ per free-fall time \citep{Faucherfeedbackregulate2013,
Hopkinsmetaldiffusion2017,OrrKSlaw2017}. We note that feedback from accreting SMBHs 
is not included in the simulations.

We analyze 14 Milky Way (MW)-mass galaxies from the public data release of 
FIRE-2 \citep{wetzelPublicDataRelease2023},
and we list them in Table~\ref{tab:sample}. 
Each of the 14 target galaxies was selected from the parent cosmological
simulation \citep{Onorbezoominsimulation2014}, and re-simulated at a higher 
resolution in a zoom-in volume enclosing the galaxy.
Eight of these, with names prefixed with {\objname m12}, 
are isolated galaxies taken from the Core suite
\citep{GarrisonDwarfsatellitesabundence2019,Garrisonsubhalo2017,Wetzeldwarfgalaxy2016,
hopkinsFIRE2SimulationsPhysics2018,SamuelFIREsatellite2020}, while the rest,
({\objname Romeo}, {\objname Juliet}, {\objname Thelma}, 
{\objname Louise}, {\objname Romulus} and {\objname Remus})
resemble the Local Group which is dominated by a pair of galaxies similar to the MW-M31 pair
\citep{GarrisonDwarfsatellitesabundence2019,GarrisonDwarfsatellitessfr2019}.
Dark matter halo properties were obtained from {\software Rockstar} 
\citep{behrooziROCKSTARPHASESPACETEMPORAL2012} by using only dark matter particles. 
Galaxies were then identified using {\software HaloAnalysis}  \citep{Wetzelhaloanalysis2020}, 
a slightly modified version of {\software Rockstar-Galaxies}, which itself 
is a version of {\software Rockstar} with support to treat particles of multi-mass and 
multi-species. Merger trees of subhalos are constructed by {\software ConsistentTrees} \citep{behrooziGravitationallyConsistentHalo2013}.
The center of a halo is defined as the location where the density reaches the 
maximum. The center of a galaxy is the center-of-mass of the member star particles. 
The virial mass of a halo, $M_{\rm vir}$ is defined as the 
total mass of all particles within the virial radius, $R_{\rm vir}$,
defined by the top-hat window in the spherical collapse model \citep{bryanStatisticalPropertiesXRay1998}. 
The virial velocity is defined as $V_{\rm vir} \equiv \sqrt{GM_{\rm vir}/R_{\rm vir}}$.
The virial density, $\bar{\rho}_{\rm vir}$, is defined as the mean matter density 
within $R_{\rm vir}$.
The mass of dark matter within $R_{\rm vir}$ is denoted as $M_{\rm vir, dm}$.
The stellar radius of a galaxy, $R_{50,*}$ ($R_{90,*}$), is computed as the radius 
enclosing $50\%$ ($90\%$) of the mass of member star particles in the galaxy
\citep{SamuelFIREsatellite2020}. For isolated galaxies, the initial mass of baryon particles 
(both gas and star) is $M_{\rm baryon} = 7070 \Msun$, with the exception that {\objname m12z} 
has $4200 \Msun$, while the Local-Group analogues have $M_{\rm baryon} = 3500$--$4000\Msun$. 
The softening length of gravity is $2.7$--$4.4\pc$ for star particles, and $31$--$40\pc$ for 
dark matter particles. The minimum adaptive force softening is adopted for gas particles, 
and it ranges from $0.4$--$1.0\pc$. In Appendix~\ref{app:sample_selection}, we compare 
the halo formation histories and environments of the 14 galaxies adopted in our analysis
with those of a large (and more complete) sample of MW-mass galaxies from 
lower-resolution simulation in a large box to show that the adopted sample
is not strongly biased in terms of halo assembly 
relative to the general population. 

The cosmology adopted by FIRE-2 is a flat $\Lambda$CDM, with parameters 
varying among zoom-in runs: the density parameters $\Omega_{\rm m}=0.26$--$0.31$, 
$\Omega_{\rm b}=0.044-0.048$, the Hubble parameter $h=0.68$--$0.71$ and 
the power-spectrum normalization $\sigma_8=0.8$--$0.82$.
The cosmic time is denoted by $t$. 

\begin{figure*}
    \begin{minipage}
    [t]{0.95\textwidth}
        \centering
        \includegraphics[width=0.95\textwidth]{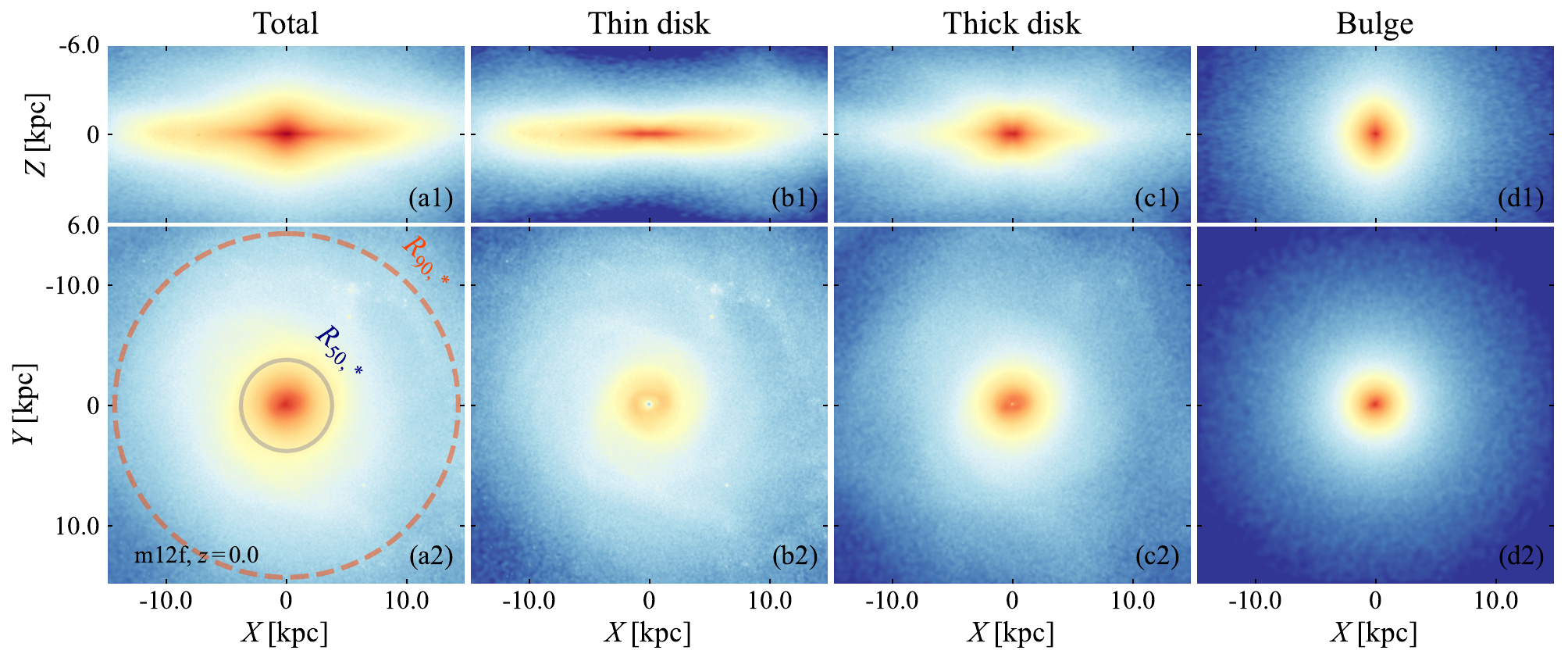}
        \label{fig:visualization1}
    \end{minipage}
    \begin{minipage}
        [t]{0.95\textwidth}
            \centering
            \includegraphics[width=0.95\textwidth]{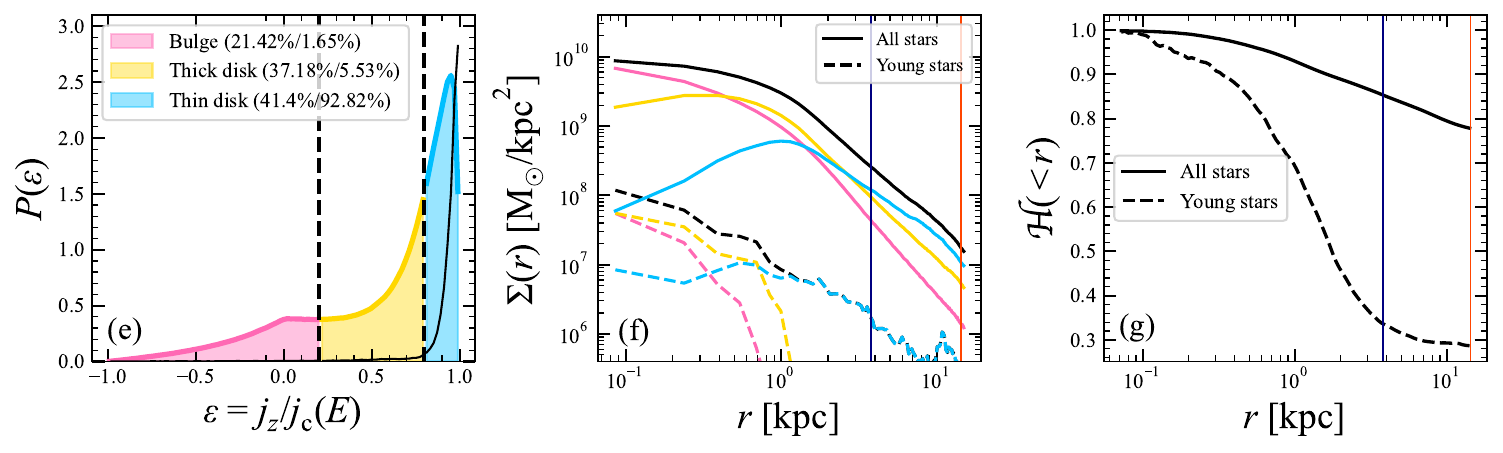}
            \label{fig:visualization2}
    \end{minipage}
    \caption{{\figem Visualization and profiles of the galaxy \texttt{m12f} at $z=0$}.
    {\figem (a)}--{\figem (d)}, 2-D stellar mass surface density, viewed edge-on (row 1) and 
    face-on (row 2). The four columns show the results for all stellar particles, 
    and those belonging thin disk ($\epsilon\geq0.8$), thick disk ($0.2\leq\epsilon<0.8$)
    and bulge ($\epsilon<0.2$), respectively.
    {\figem (e)}, mass-weighted distribution of circularity ($\epsilon$) of 
    star particles. Colored solid curve shows the distribution of all star particles 
    within $R_{\rm 90,*}$, with three colors indicating the three kinematic 
    components, respectively. Black solid curve shows the distribution of 
    only young star particles (age $<100\Myr$) within $R_{\rm 90,*}$.
    The mass fraction of each component in indicated in the legend for all/young stars.
    {\figem (f)}, mass surface density profiles, obtained using all (solid curves) 
    and young (dashed curves) stars, showing in total (black) and separately for
    the three kinematic components (colored).
    {\figem (g)}, hotness as a function of radius, obtained by using all 
    (solid curve) and young (dashed curve) stars enclosed within the radius.
    We mark the galaxy sizes, $R_{\rm 50,*}$ (blue) and $R_{\rm 90,*}$ (red), 
    by circles in {\figem (a2)} and vertical lines in 
    {\figem (f)} and {\figem (g)}. 
    }
    \label{fig:visualization}
\end{figure*}

\begin{figure*}
	\includegraphics[width=0.95\textwidth]{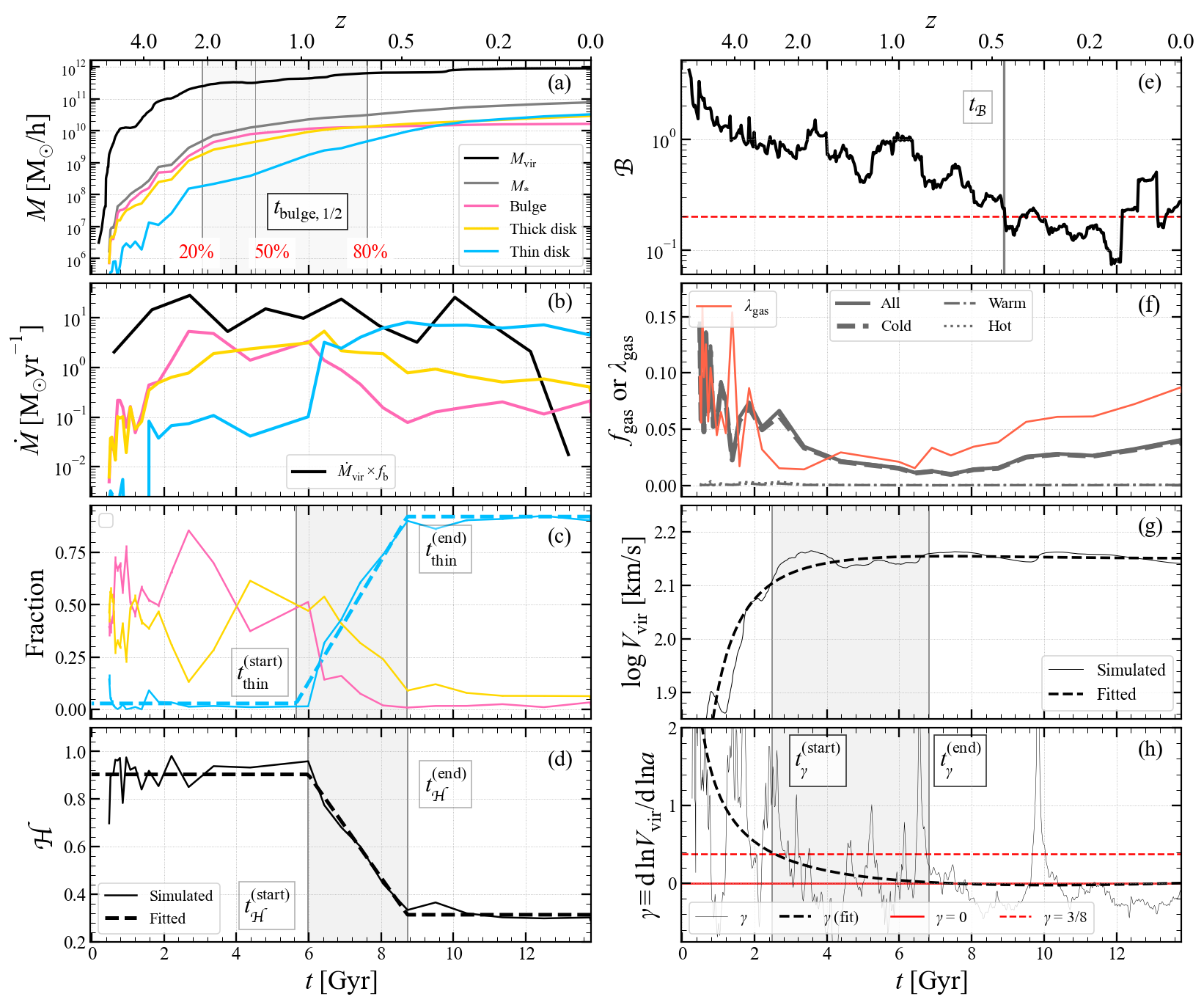}
    \caption{ {\figem Properties of the galaxy \texttt{m12f}
    and its host halo as functions of the cosmic time ($t$) or redshift ($z$).}
    Each panel shows a different property.
    {\figem (a)}, stellar mass in total (gray) and separately 
    for the three kinematic components (colored),
    and $M_{\rm vir}$ of the host halo (black).
    Three vertical lines indicate the epochs at which the bulge mass
    reaches 20\%, 50\% and 80\%, respectively, of its final value at $z=0$,
    with that corresponding to 50\% denoted as $t_{\rm bulge, 1/2}$.
    {\figem (b)}, the growth rate of the three kinematic components (colored).
    For reference, we show the accretion rate of baryons, defined as
    the growth rate of $M_{\rm vir}$ times the cosmic baryon fraction (black).
    {\figem (c)}, mass fractions of the three kinematic components, 
    using young stars only. Grey vertical band indicates the transitional 
    interval, $t_{\rm thin}^{\rm (start)}$--$t_{\rm thin}^{\rm (end)}$, 
    of the thin-disk fraction.
    {\figem (d)}, dynamical hotness of the galaxy ($\mathcal{H}$), calculated using
    young stars only. 
    Grey vertical band indicates the transitional interval,
    $t_{\mathcal{H}}^{\rm (start)}$--$t_{\mathcal{H}}^{\rm (end)}$
    of $\mathcal{H}$.
    {\figem (e)}, burstiness of star formation ($\mathcal{B}$). Vertical line 
    indicates $t_{\mathcal{B}}$, the epoch when the star formation transits 
    from bursty to steady (defined by $\mathcal{B} = 0.2$, indicated by the dashed
    horizontal line).
    {\figem (f)}, cold-gas spin $\lambda_\text{cold-gas}$,
    and fractions of cold ($f_\text{cold-gas}$), warm and hot gas.
    {\figem (g)}, {\figem (h)}, 
    virial velocity ($V_{\rm vir}$) and specific growth rate of the host halo 
    ($\gamma$; see Eq.~\ref{eq:halo-gamma} for definition).
    Grey vertical bands indicate the transitional interval,
    $t_{\gamma}^{\rm (start)}$--$t_{\gamma}^{\rm (end)}$,
    of halo assembly, defined by $\gamma = 3/8$--$0$ 
    (horizontal lines in {\figem h})
    In {\figem (c)}, {\figem (d)}, {\figem (g)} and {\figem (h)}, each 
    solid curve is obtained from the simulation, while the associated dashed curve
    is obtained by a parametric fitting (see \S\ref{ssec:galaxy-properties}
    and \S\ref{ssec:halos} for details).
    This figure shows that the galaxy evolves
    from an early phase featured by hot dynamics, bulge-dominated structure 
    and bursty star formation, to a later phase featured by cold dynamics,
    disk-dominated structure and steady star formation, and that
    the assembly of the host halo decelerates with time.
    See \S\ref{sec:co-evolution} for details.
    }
    \label{fig:Example}
\end{figure*}

\subsection{Properties of the simulated galaxies} 
\label{ssec:galaxy-properties}

{\renewcommand{\arraystretch}{1.2}
\begin{table}
\caption{{\figem Properties of galaxies and halos defined and analyzed in this paper.}}
\label{tab:properties}
\centering
\tabcolsep=0.1cm
\begin{tabular}{>{\centering\arraybackslash}m{4em}|m{24em}}
    \hline
    \makecell{\bf Property} & \makecell{\bf Description} \\
    \hline
    $\epsilon$ & Orbital circularity of a particle (\S\ref{sssec:decomposition}) \\
    $f_{\rm thin}$ & Mass fraction of thin disk (\S\ref{sssec:decomposition}); similar 
        definitions are made for thick disk ($f_{\rm thick}$) and bulge ($f_{\rm bulge}$) \\
    $\mathcal{H}$ & Dynamical hotness (\S\ref{sssec:hotness-def})\\
    $\mathcal{B}$ & Burstiness of the star formation (\S\ref{sssec:burstiness})\\
    $n$ & S\'ersic index of bulge (\S\ref{sssec:pros-of-bulge})\\
    $c/a$ & Minor to major axis ratio of bulge (\S\ref{sssec:pros-of-bulge}); similar 
        definition is made for intermediate to major axis ratio ($b/a$) \\
    $T$ & Triaxiality of bulge (\S\ref{sssec:pros-of-bulge})\\
    $f_\lambda$ & Ratio between cold-gas fraction and cold-gas spin (\S\ref{sssec:f-gas}) \\
    $f_{\text{gas+star}}$ & Mass fraction of cold gas plus young stars within some radius (\S\ref{sssec:f-gas}) \\
    $\gamma$ & Specific growth rate of halo (\S\ref{ssec:halos}) \\
    \hline
    \end{tabular}
\end{table}
}

\subsubsection{Stellar kinematics and decomposition}
\label{sssec:decomposition}

Kinematic decomposition in this paper is performed using a circularity-based method
similar to that of \citet{yuBurstyOriginMilky2021,
yuBornThisWay2023}. For each galaxy at a given snapshot, we
categorize each of its member star particles into one of the three components:
bulge, thick disk and thin disk, according to the circularity \citep{Abadicircularity2023}
of the star particle: 
\begin{equation}
    \epsilon = j_z/j_{\rm c}(E)   \,,
\end{equation}
where $j_z$ is the angular momentum of that particle projected to 
the direction of $\bm{J}_{\rm tot}$, the net angular momentum of all stellar particles within 
$R_{\rm 90,*}$ at that snapshot \citep{gurvichRapidDiscSettling2023},
$j_{\rm c}(E)$ is the angular momentum of a circular orbit with the same 
energy ($E$) as that of the star particle.
Particles with 
$-1 \leqslant \epsilon < 0.2$,
$0.2 \leqslant \epsilon < 0.8$
and $0.8 \leqslant \epsilon \leqslant 1$
are categorized as bulge, thick disk and thin disk, respectively.
Previous studies adopted more rigorous kinematic classification methods 
\citep[e.g.][]{duIdentifyingKinematicStructures2019,duKinematicDecompositionIllustrisTNG2020,zanaMorphologicalDecompositionTNG502022,Liangdarkhalogalaxy2025}, 
such as classifications in the plane of angular momentum versus orbital energy. 
However, our work focuses primarily on morphological transitions, particularly in the 
evolution of the significance of the disk component, which can be 
characterized robustly using the angular momentum alone. 
We therefore employ a simplified kinematic decomposition based solely on the angular 
momentum to define the three components. We define the direction 
of $\bm{J}_{\rm tot}$ as the normal direction (the $z$ axis) of the disk 
($x$-$y$) plane. The projections of a simulated galaxy perpendicular to and along 
the $z$-axis are defined as the edge-on and face-on views of the galaxy, respectively.

Fig.~\ref{fig:visualization} shows an example galaxy, {\objname m12f}, at $z = 0$,
with the three components shown separately and in total, viewed edge-on and face-on. 
Here, we include all particles within $R_{\rm 90,*}$. Panel (e) shows the distribution 
of $\epsilon$ for all stellar particles within $R_{\rm 90,*}$, weighted by their 
masses. Panel (f) shows the face-on surface density profile of the three components and of the total, using 
all particles within $R_{\rm 90,*}$. The thin disk is extended, dominates the surface density 
in the outer region, and follows an exponential profile at $r \gtrsim 3\Kpc \approx R_{\rm 50,*}$. 
In the inner region ($r\lesssim 1\Kpc$), however, the thin disk is 
truncated. This truncation is similar to that found in other simulations 
(\citealt{duKinematicDecompositionIllustrisTNG2020}, see their fig. 1; 
\citealt{buckStarsBarsII2019}, see their fig. 2) and in observations 
with spatially-resolved spectroscopy (\citealt{zhuOrbitalDecompositionCALIFA2018}, 
see their fig. 8 and \S4), and suggests a potential caveat of 
using an un-truncated profile to fit the entire cold-disk component.
On the other hand, the thick disc is more vertically extended,
reflecting its hotter dynamics with a higher degree of random-motion 
support for the vertical structure. Both the images (c1, c2) and the surface density 
profile (yellow in f) suggest the existence of an inner, concentrated component at 
around $1\Kpc$, in alignment with the concept of slow-rotating ``pseudo-bulge'' 
found in observations \citep[e.g.][]{KormendyPesudobulgeKormendy,
fisherObservationalGuideIdentifying2016,
huResolvedPropertiesClassical2024}
and in hydrodynamic simulations \citep[e.g.][]{okamotoOriginPseudobulgesCosmological2013,
gargiuloPrevalencePseudobulgesAuriga2019,
duKinematicDecompositionIllustrisTNG2020,
zanaMorphologicalDecompositionTNG502022}. 
The bulge dominates the inner $0.5\Kpc$, but it also has a faint extension
to the outer region ($\gtrsim 10\Kpc$), consistent with being a combination of 
the ``classical bulge'' and the stellar halo found  
in observations
\citep[e.g.][]{KormendyPesudobulgeKormendy,fisherObservationalGuideIdentifying2016,
huangDetectionEnvironmentalDependence2018,huResolvedPropertiesClassical2024}
and in hydrodynamic simulations
\citep[e.g.][]{duKinematicDecompositionIllustrisTNG2020,zanaMorphologicalDecompositionTNG502022}.

The three components have similar mass within $R_{\rm 90,*}$, as can be 
seen from the mass-weighted distribution of $\epsilon$ (panel e),  
consistent with the results of an earlier analysis of the FIRE-2 by \cite{yuBurstyOriginMilky2021}. 
All these show that FIRE-2 is able to produce realistic kinematics of 
MW-size galaxies, with components similar to those found in observations.

To quantify the kinematics of star-forming gas within a target galaxy, we 
utilize in-situ young stars, defined as bound stellar particles with ages 
$\leq 100\Myr$ and birthplaces within $\leq R_{90,*}$ from the center of the galaxy. 
This selection is made to avoid complexity introduced by potential post-formation 
heating of old stars and the contamination by ex-situ stars.
Hereafter, we use these in-situ young stars to define transition times in 
the morphological type of galaxy. As shown in Fig.~\ref{fig:visualization}(e), the 
distribution of $\epsilon$ of in-situ young stars 
at $z=0$ is sharply peaked around $\epsilon = 1$.
Most ($\approx 93\%$) of the in-situ young stars at this epoch belong to the thin disk, 
and only a small fraction ($\approx 1.6\%$) is categorized into the bulge component.
Fig.~\ref{fig:visualization}(f) reinforces this conclusion by comparing the face-on
surface densities of in-situ young stars (dashed lines) with those of
all stars (solid lines), which shows that in-situ young stars are mostly
contained in the extended thin disk.

Fig.~\ref{fig:Example}(a) shows the stellar mass as a function of
the cosmic time ($t$), for {\objname m12f} as a total and separately 
for the three components. Here, we first measure the radius that encloses 90\% of stellar mass 
of the corresponding component and then estimate the stellar mass within this radius.  
For reference, we also include the mass of the host 
halo, $M_{\rm vir}$. In parallel, panels (b) and (c) show the corresponding 
mass growth rate, and mass fraction of in-situ young stars in different kinematic 
components, respectively. Remarkably, at $z \gtrsim 1$, stars are mostly born in 
dynamically hot (bulge and thick-disk) components, and the fraction in the cold, 
thin disk component is very small. The mass fractions in the bulge and thick-disk
components are comparable and show similar redshift dependence.
At $z \approx 1$, the star-formation activity rapidly migrates to the thin 
disk, outpacing that in the hot components within a timescale of 
$\approx 1\Gyr$. Thereafter, the mass in the bulge changes little, while 
that in the thin disk increases steadily. The behavior of 
$M_{\rm bulge}$ versus $t$ prompts the definition of a formation 
time, $t_{\rm bulge,1/2}$, which we set to be the time when the bulge has
reached half ($50\%$) of its final mass and is marked by a vertical line in panel (a). 
For reference, we also mark the times for the formation of $20\%$ and $80\%$ of 
the final bulge mass.  

The transition of star formation from the dynamically hot to cold
components shown in Fig.~\ref{fig:Example}(c) motivates us to define 
a transition time to separate the early, hot phase from the late, cold phase, 
for each individual galaxy. To do this, we fit the fraction of young stars 
in the thin-disk component ($f_{\rm thin}$) as 
a function of cosmic time ($t$), as shown by the blue solid curve in 
Fig.~\ref{fig:Example}(c) for example, by the following piecewise function:
\begin{equation}
f_{\rm thin} = 
\begin{cases}
    f_1                                                            \,,& t < t_1; \\
    \frac{t - t_1}{t_2-t_1} ( f_2-f_1 ) + f_1                       \,,\ \ \ \ \ \ \ \ & t_1 \leqslant t< t_2 \,; \\ 
    f_2                                                             \,,& t\geqslant t_2 \,. \\ 
\end{cases}
\label{eq:fitting}
\end{equation}
Here, $f_1$ and $f_2$ are the initial and final values of $f_{\rm thin}$, respectively,
introduced to describe the shape of $f_{\rm thin}$ seen in 
Fig.~\ref{fig:Example}(c) at $t < t_1$ and $t \geqslant t_2$.
The change of $f_{\rm thin}$ in the transitional interval 
between $t_1$ and $t_2$ is described by a linear function.
The fitting result for \texttt{m12f} is shown by the blue dashed curve in
Fig.~\ref{fig:Example}(c), with the duration of the transitional phase 
($z \approx 0.5$--$1$, or $t\approx 5.5$--$8.5\Gyr$)
indicated by the gray shade. To avoid ambiguity, we will denote $t_1$ and $t_2$ as 
$t_{\rm thin}^{\rm (start)}$ and $t_{\rm thin}^{\rm (end)}$, respectively.
We use the mean value of the two,  
\begin{equation}
 t_{\rm thin} = \frac{1}{2} \left[t_{\rm thin}^{\rm (start)}+t_{\rm thin}^{\rm (end)}\right] \,,
\end{equation}
to represent the epoch of the morphology transition, and
use the interval [$t_{\rm thin}^{\rm (start)}$, $t_{\rm thin}^{\rm (end)}$] 
to bracket the uncertainty. For comparison, we denote the fraction in
the thick-disk and bulge components as $f_{\rm thick}$ and $f_{\rm bulge}$, 
respectively.

\subsubsection{Dynamical hotness and phases of galaxies}
\label{sssec:hotness-def}

To describe the transition behavior of FIRE-2 galaxies while avoiding
arbitrariness in the threshold separating the bulge and disk components, 
we define another physical parameter, {\em hotness}, as the ratio 
between the stellar velocity dispersion and the total velocity within some radius, $R$:
\begin{equation}
    \mathcal{H} = \frac{\sigma_{\rm v}}{\left| \bm v \right|} \,.
\label{eq:def-hotness}
\end{equation}
Here, the total velocity is computed as the mass-weighted total velocity
of star particles:
\begin{equation}
    \left| \bm v \right| = \sqrt{\frac{ 
    \sum_{i} M_i {\bm v}_i^2    
}{\sum_i M_i}} \,,
\end{equation}
where $M_i$ and ${\bm v}_i$ are the mass and velocity vector, respectively,
of the $i$-th star particle, and the summation is over 
all in-situ young star particles 
within $R$ that are bound to the galaxy.

To remove rotational motion from the computation of 
$\sigma_{\rm v}$, we find the azimuthal velocity, $v_\phi$, of each star 
particle in the cylindrical coordinate system (centered on the galaxy center, 
with the cylindrical axis parallel to $\bm{J}_{\rm tot}$).
We obtain the radial profile, $v_\phi(r_{\rm p})$, as the mass-weighted $v_\phi$ 
of the star particles in bins of cylindrical radius, $r_{\rm p}$.
We define the rotation-subtracted velocity of the $i$-th star particle as
${\bm v}'_i \equiv {\bm v}_i - v_\phi (r_{\rm p}) \hat{{\bm e}}_{\phi,i}$,
with $\hat{{\bm e}}_{\phi,i}$ being the unit vector in the azimuthal direction at the 
location of the star particle.
The rotation-subtracted 3-D velocity dispersion is then computed as
\begin{equation}
    \sigma_{\rm v} = \sqrt{\frac{ 
        \sum_i M_i \left|  {\bm v}'_i -  \left<{\bm v}'\right>  \right|^2 
        }{ \sum_i M_i}} \,,
\end{equation}
where $\left<{\bm v}'\right>$ is the mass-weighted mean value of 
the rotation-subtracted velocities of star particles. 
The calculations of $v_\phi(r_{\rm p})$, $\left<{\bm v}'\right>$ and 
$\sigma_{\rm v}$ are performed by using the same set of star particles as used for $\left| \bm v \right|$.
So defined, the hotness $\mathcal{H}$ is a dimensionless quantity, whose square
is the fraction of kinetic energy in the form of random motion. 
A value of $\mathcal{H} \sim 1$ thus corresponds to a dynamically 
hot system supported purely by random motion, usually a bulge, while
$\mathcal{H} = 0$ corresponds to a dynamically cold system supported
purely by rotation, usually a thin disk.

Fig.~\ref{fig:visualization}(g) shows the hotness as a function of 
radius for {\objname m12f} at $z=0$ for all stars (solid) and
in-situ young stars (dashed).
For all stars, $\mathcal{H}$ decreases monotonically with radius.
The inner region ($R \lesssim 0.3\Kpc$) is dispersion-dominated, 
with hotness $\mathcal{H} \approx 1$.
In contrast, on large radii ($R \approx 10 \Kpc$),
$\mathcal{H} \approx 0.8$, indicating that random motion is 
still significant in the outer region, but less dominating compared to that in
the inner region.
The in-situ young stars have significantly lower hotness, consistent with
their lower fraction in hot components (bulge and thick disk),
as shown in Fig.~\ref{fig:visualization}(e), (f). 
Prominently, $\mathcal{H}$ of in-situ young stars on large radii of
$R \gtrsim 3\Kpc$ is as low as about $0.3$,
indicating that most of the kinetic energy is in the form of ordered rotation,
consistent with the dominance of the thin-disk component there (panel f).
All these findings support the hotness $\mathcal{H}$ as a viable parameter
to quantify the dynamical state of a galaxy within a given radius. 
In the following, we use $\mathcal{H}$ within $R = R_{\rm 90,*}$ to represent 
the hotness of a galaxy as a whole.


Fig.~\ref{fig:Example}(d) shows the evolution of the hotness for {\objname m12f}.
The value of $\mathcal{H}$ remains at a constant value of $0.9$ 
at $z \gtrsim 1$, rapidly decreases during the period of $z \approx 0.5$--$1$, 
and remains $\approx 0.3$ thereafter. The steps at both high and low $z$,
as well as the rapid decrease between them, are similar to those seen 
in $f_{\rm thin}$ described in \S\ref{sssec:decomposition} and shown in 
Fig.~\ref{fig:Example}(c). We therefore use the same piecewise function form 
as given by Eq.~\eqref{eq:fitting} to fit the hotness $\mathcal{H}$ as a function of
time. We define the starting and ending times of the 
transitional stage as $t_{\ssc \mathcal{H}}^{\rm (start)}$ and 
$t_{\ssc \mathcal{H}}^{\rm (end)}$, respectively, and use $t_{\ssc \mathcal{H}}$ 
to denote the mean of them.
Dashed curve in Fig.~\ref{fig:Example}(d) shows the fitting result, together 
with the two transition times. Because of the similarity between 
$\mathcal{H}(t)$ and $f_{\rm thin}(t)$, the two transition times 
are also about the same between the two functions.

Based on the shape of $\mathcal{H} (t)$ and the corresponding change 
in the galaxy kinematics, we define, for each galaxy, the following 
dynamical phases:
\begin{enumerate}[topsep=0pt,parsep=0pt,itemsep=0pt]
    \item {\bf\em The hot phase}, 
    defined by the time interval 
    $t < t_{\ssc \mathcal{H}}^{\rm (start)}$, in which 
    $\mathcal{H}$ is constantly high;
    \item 
    {\bf\em The cold phase}, defined by the time interval 
    $t \geqslant t_{\ssc \mathcal{H}}^{\rm (end)}$, in which the hotness 
    is constantly low;
    \item 
    {\bf\em The transitional phase}, defined by the time interval 
    $t_{\ssc \mathcal{H}}^{\rm (start)} \leqslant t < t_{\ssc \mathcal{H}}^{\rm (end)}$, 
    in which the hotness decreases rapidly with time.    
\end{enumerate}
Note that such an evolution pattern was also found by 
\citealt{mccluskeyDiscSettlingDynamical2024}, who focused on 
disk settling. The terms `Pre-Disc',  `Early-Disc' and 
`Late-Disc' they introduced are similar to our definitions
of different phases based on $f_{\rm thin}(t)$ instead of $\mathcal{H}(t)$.

\subsubsection{Burstiness in star formation}
\label{sssec:burstiness}

MW-size galaxies in FIRE-2 are found to show bursty (steady) star-formation history in the 
early (late) time, with transition of the `burstiness' coinciding with
the dynamical transition \citep{sternMaximumAccretionRate2020,
FloresSFRburstyFIRE2021,yuBurstyOriginMilky2021,yuBornThisWay2023,
hopkinsWhatCausesFormation2023}.
Following \citealt{yuBurstyOriginMilky2021}, we define the burstiness of star
formation, $\mathcal{B}$, as the ratio of the standard deviation of 
the instantaneous star formation rate (SFR) within a time window 
$\Delta t = 500\Myr$ to the averaged SFR within the same window:
\begin{equation}
    \mathcal{B} = \frac{ \sqrt{{\rm Var}_{\Delta t}\left[{\rm SFR}\right]} }{ \left<{\rm SFR}\right>_{\Delta t}} \,.
\end{equation}
Here, the instantaneous SFR is estimated by the mass of stars formed
per unit time during a shorter time interval of $10\Myr$
within $10\Kpc$ from the center of the most massive progenitor
at the time $t$ considered,
and the variance, `Var', and average, `$\left<\right>$',
are computed over $[t-\Delta t/2, t+\Delta t/2]$.
So defined, a value of $\mathcal{B} \gtrsim 1$ indicates a highly bursty
star formation, while $\mathcal{B} \ll 1$ indicates that the star formation is steady. 
Fig.~\ref{fig:Example}(e) shows the evolution of $\mathcal{B}$ for {\objname m12f}.
As can be seen, $\mathcal{B}$ decreases from $\approx 1$ at $z = 4$ to $\approx 0.2$ at $z=0$. 
A sudden increase in $t \gtrsim 12\Gyr$ can be observed, 
which is induced by a prograde merger of LMC size \citep[e.g.][]{yuBurstyOriginMilky2021}. 
However, the hotness $\mathcal{H}$ of the entire galaxy appears to be
stable during this merger, reflecting the limited impact of individual
minor mergers on the hotness of the galaxy. Similarly to \citet{yuBurstyOriginMilky2021},
we define the transition time in the burst of star formation, $t_{\ssc\mathcal{B}}$,
as the time when $\mathcal{B}$ first falls below 0.2. For {\objname m12f}, 
$t_{\ssc\mathcal{B}} \approx 9\Gyr$, as shown by the vertical line in Fig.~\ref{fig:Example}(e). 

\subsubsection{Structure of stellar bulge}
\label{sssec:pros-of-bulge}

A particular focus of this paper is the origin and evolution of dynamically 
hot stellar components (bulges) of galaxies. In the following, we 
define a number of parameters to quantify the structural properties of stellar bulges.
The results will be presented in \S\ref{sec:early-growth}.

To quantify the anisotropy of stellar distribution in a bulge,
we approximate the bulge by an ellipsoid whose orientation 
and axis lengths are defined, respectively, by the eigenvectors and eigenvalues of the 
weighted inertia tensor, 
\begin{equation}
\mathcal{I}_{i j} \equiv  \frac{ \sum_nm_{n} x_{n, i} x_{n, j}}{\sum_n m_{n} r_n^2}\,.
\end{equation}
Here $i,\,j = 1,\,2,\,3$ are the three spatial dimensions;
$m_n$ is the mass of the $n$-th member star particle,
and $x_{n,i}$ and $x_{n,j}$ are its galactocentric coordinates in the 
corresponding dimensions. The elliptical distance of the stellar particle,
$r_n$, is defined as
\begin{equation}
r_n=\sqrt{x_{n,1}^2 + \frac{x_{n,2}^2 }{q^2} + \frac{x_{n,3}^2}{s^2}}\,,
\end{equation}
where  $q \equiv b/a$ and $s\equiv c/a$ are axis ratios;
$a$, $b$, $c$ $\propto$ $\sqrt{\lambda_1}$, $\sqrt{\lambda_2}$, $\sqrt{\lambda_3}$ are 
the lengths of the three axes, respectively;
$\lambda_1 \geqslant \lambda_2 \geqslant \lambda_3$ are the 
eigenvalues of the inertia tensor $\mathcal{I}_{i j}$.
The weighting $r_n^{-2}$ is adopted to avoid over-weighting particles
at large radii, as introduced by \citet{Allgoodhaloshape2006}. The summation is over all star 
particles within $R_{90,*}$, and the calculations are performed 
only when the number of particles is greater than 1,000.
The eigenvalue-dependence in the definition of the inertia tensor 
leads to a dependency loop. We thus start from a sphere (i.e. $q = 1$ and $s = 1$),  
iteratively compute the inertia tensor and obtain its eigenvectors,
and rotate the coordinates $(x_{n, 1}, x_{n, 2}, x_{n, 3})$ to the 
frame defined by the eigenvectors, until convergence is reached.
Finally, we define the triaxiality parameter,
\begin{equation}
    T \equiv \frac{1-q^2}{1-s^2} = \frac{a^2-b^2}{a^2-c^2}\,,
\end{equation}
to indicate the anisotropy in the spatial distribution of stars.
So defined, a perfect prolate ellipsoid has $T = 1$, a perfect oblate 
ellipsoid has $T = 0$, and other cases all have $0 < T < 1$.

To quantify the concentration of the stellar distribution in the bulge,
we project all bulge stars within $R_{\rm 90,*}$ along ${\bm J}_{\rm tot}$, 
measure the surface density profile as a function of the projected 
galactocentric distance, $r_{\rm p}$, and fit the profile using a S\'ersic function:
\begin{equation}
    \Sigma_*(r_{\rm p})= \Sigma_{\rm e,*} \exp \left\{
        -b_{\rm s}\left[\left(\frac{r_{\rm p}}{R_{\rm e,*}}\right)^{1 / n}-1\right]\right\}\,,
\end{equation}
where $R_{\rm e,*}$ is the effective radius, 
$\Sigma_{\rm e,*}$ is the surface density at that radius,
$n$ is the S\'ersic index that sets the concentration
of the profile, and $b_{\rm s}$ is a parameter introduced to
support the meaning of other parameters.
The fitting is done by minimizing the
$\chi^2$ loss function between the function and the simulated profile:
\begin{equation}
    \chi^2(\Sigma_{\rm e,*}, b_{\rm s}, R_{\rm e,*}, n)
    =\sum_{i} \frac{
        \left[M_i-\hat{M}_{i}(\Sigma_{\rm e,*}, b_{\rm s}, R_{\rm e,*}, n)\right]^2}{M_i}
    \,, \label{eq:chi2-Sersic}
\end{equation}
where $M_i$ and $\hat{M}_i$ are the stellar masses obtained from
the simulation and expected from the fitting function, respectively, in 
the $i$-th bin of $r_{\rm p}$.

\subsubsection{The cold-gas fraction and internal instability}
\label{sssec:f-gas}

The transitional features in the morphology, dynamical hotness and star-formation
burstiness suggest a star-formation environment in the early universe that 
is substantially different from that in the local universe. To gain insight into
the physical origin of the transition, here we follow the `quadrant' diagram 
introduced by \citetalias{moTwophaseModelGalaxy2024} (see their \S3.3) to consider 
two factors, one internal and the other external. 
The first factor, introduced in the following, is related to the cold gas 
fraction in a galaxy that determines the instability of the gas, 
while the second, related to the assembly of dark matter halos, is introduced 
in \S\ref{ssec:halos}. We will use both to understand the evolution of dynamical hotness 
and transition  of galaxy properties in \S\ref{sec:driving-processes}.

To quantify the internal instability caused by the self-gravity of a 
gaseous cloud or a disk against the support of angular momentum, we
use the ratio between the cold-gas fraction and the gas spin parameter:
\begin{equation}
    f_{\lambda} = \frac{f_{\rm gas}}{\lambda_{\rm gas}} \,,
\end{equation}
where $f_{\rm gas}$ is the ratio of the mass of cold gas
(with temperature $T < 2\times 10^4\Kelvin$) within $R_{\rm 90,*}$ to the virial
mass of the dark matter halo ($M_{\rm vir}$), 
$\lambda_{\rm gas}=|{\bm j}_{\rm gas}|/\left( \sqrt{2} V_{\rm vir} R_{\rm vir} \right)$,
with ${\bm j}_{\rm gas}$ being the specific angular momentum of the cold gas within $R_{\rm 90,*}$. 
Fig.~\ref{fig:Example}(f) shows $f_{\rm gas}$ and $\lambda_{\rm gas}$ versus $t$ for {\objname m12f}. 
This example shows that the galaxy starts from an early phase where both $f_{\rm gas}$ and $\lambda_{\rm gas}$
are fluctuating around relatively high values. From $z \approx 3$ to $\approx 1.5$, 
$f_{\rm gas}$ is much higher than $\lambda_{\rm gas}$, and at $z<1$ the order  
reverses. 

The ratio $f_\lambda$ is defined by \citet[see their \S3]{moFormationGalacticDiscs1998}
to describe the rotational support of a gaseous disk against self-gravity,
with the assumption that the gaseous disk is embedded in a dark matter halo
characterized by a static NFW profile that dominates the gravity of all 
collisionless components.
Here we utilize the spin of cold gas, instead of that of dark matter,
to account for the potential difference between the two
\citep{jiangDarkmatterHaloSpin2019,Yanggalaxysizehalospin2023}. 
Depending on the concentration of the halo, the baryon fraction and distribution,
the threshold for the gaseous disk to be stable against self-gravity varies around $1$
\citep[e.g.][]{Efstathioudiscstability1982,moFormationGalacticDiscs1998}. 
Thus, $f_{\lambda} \lesssim 1$ ($\gtrsim 1$) 
implies a stable (unstable) disk. In \S\ref{sec:driving-processes}, we will use
the evolution of $f_{\rm gas}$, $\lambda_{\rm gas}$ and $f_\lambda$ to understand the 
formation of different components of galaxies.

The ratio $f_{\lambda}$ defined above is a condition for the gas to become self-gravitating
before it can be supported by angular momentum. To study how gas collapses produce
a self-gravitating structure near the halo center, we define a parameter, 
\begin{equation}
    f_{\text{gas+star}} = \frac{M_{\rm gas}+M_{\text{young-star}}}{M_{\rm tot}} \,,
\end{equation}
where $M_{\rm gas}$ is the mass of cold gas, $M_{\text{young-star}}$ is the mass of stars 
formed in the past duration of one dynamical timescale of the host halo, 
$t_{\mathrm{dyn,h}}(t)\equiv\sqrt{(3 \pi) /(16 G {\overline\rho}_{\rm vir}(t))}$,
and $M_{\rm tot}$ is the total mass (dark matter plus baryon mass). 
All the masses here should be evaluated within a common aperture, which 
we will specify when we use this parameter.
The inclusion of young stars in $f_{\text{gas+star}}$ is to take into account 
the fact that processes on galactic scales have timescales typically of the dynamical
timescale and that some star-forming gas may have already been turned into stars 
in this duration.


\subsection{Dark matter halos hosting the galaxies}
\label{ssec:halos}

The assembly history of a dark matter halo was found to have two phases
according to the growth rate \citep{zhaoGrowthStructureDark2003}.
Following \citetalias{moTwophaseModelGalaxy2024} (see their \S2), we 
define the specific growth rate of the virial velocity for a 
halo at redshift $z$, as 
\begin{equation}
    \gamma \equiv \frac{\dot{V}_{\rm vir}}{H(z) V_{\rm vir}} 
    = \dv{\ln V_{\rm vir}}{\ln a} \,,
    \label{eq:halo-gamma}
\end{equation} 
where $H(z)$ is the Hubble parameter at $z$, $a = 1/(1+z)$ is the scale factor,
and the differentiation is numerically evaluated along the main branch of the subhalo merger tree.
The black solid curve in Fig.~\ref{fig:Example}(g) shows the evolution of $V_{\rm vir}$ 
for {\objname m12f}, and that in Fig.~\ref{fig:Example}(h) shows the evolution of $\gamma$ obtained 
numerically from $V_{\rm vir}$.

The accretion of a halo, as described by $\gamma$, is often noisy, exhibiting
significant small-scale fluctuations. Part of the fluctuations may 
be produced by numerical defects in the halo finder and tree builder. 
To extract a clean accretion history that perseveres the long-term trend, 
we follow \citetalias{moTwophaseModelGalaxy2024} to fit the mass assembly history 
of each halo by a smooth function of redshift:
\begin{equation}
    \ln M_{\rm vir}(z)= c_0 + c_1 \frac{z}{1+z} + c_2 \ln (1+z)+c_3 z\,,
\label{eq:fit-M-vir}
\end{equation}
where ($c_0,c_1,c_2,c_3$) are four fitting parameters.  
The smoothed values of other halo properties, such as virial 
radius $R_{\rm vir}$, virial velocity $V_{\rm vir}$, and specific 
growth rate $\gamma$, are derived from the smoothed values of $M_{\rm vir}$.
The evolutions of smoothed $V_{\rm vir}$ and $\gamma$ obtained  
for {\objname m12f} are shown by the black dashed curves in 
Fig.~\ref{fig:Example}(g) and (h), respectively. 
A clear deceleration in the halo accretion, as represented by the evolution of $\gamma$,  
can be seen. With this smooth fitting, the assembly of a halo can be 
divided into two phases: an early phase with fast accretion (named `fast phase'), 
and a late phase with slow accretion (named `slow phase').
The threshold for such a division falls between a range of 
$\gamma_{\rm f} \approx 0$--$3/8$ 
\citep{zhaoGrowthStructureDark2003,moreSplashbackRadiusPhysical2015,bocoTwOParametersSemi2023}.
Here we use both values to bracket the uncertainty in the division. 
Specifically, we define transition times, $t_\gamma^{\rm (start)}$ and 
$t_\gamma^{\rm (end)}$, as the cosmic times when the smoothed $\gamma(z)$ first
reaches the threshold $\gamma_{\rm f} = 3/8$ and $0$, respectively,
and we denote the average of the two transition times as $t_\gamma$.
In Table~\ref{tab:sample}, we list the transition times for all halos 
in our sample. The vertical lines in Fig.~\ref{fig:Example}(g) and (h) 
bracket the range from $t_\gamma^{\rm (start)}$ to $t_\gamma^{\rm (end)}$ 
for {\objname m12f}.


For reference, we also compute the `half-mass' formation time, $t_{\rm h, 1/2}$, 
for each halo in our sample. This formation time is obtained by following the evolution of 
$M_{\rm vir}$ along the main branch of the subhalo merger tree rooted in the halo at $z = 0$, 
and solving the equation
\begin{equation}
    M_{\rm vir}(t_{\rm h, 1/2}) = \frac{1}{2} M_{\rm vir}(z=0) \,.
\end{equation}
The exact formation redshift is evaluated by linearly interpolating the redshift 
between the two adjacent snapshots where $M_{\rm vir}$ first crosses the half of 
its final mass. This time has been commonly used in the literature to
characterize the assembly of dark matter halos
\citep[e.g.][]{gaoAgeDependenceHalo2005,liHaloFormationTimes2008,lyuHalosGalaxiesVII2023,lyuHalosGalaxiesIX2024}, 
and was found to correlate with $t_\gamma$ (see Appendix B of \citetalias{moTwophaseModelGalaxy2024}). 
Another related formation time, which is also commonly used in the literature to 
characterize halo assembly, is $t_{\rm h,Vpeak}$, defined as the cosmic time when 
the maximum circular velocity ($V_{\rm max}$) of the halo 
reaches its peak along the main-branch history. This time was found to be 
sensitive to violent changes in the halo mass, and has been used to 
incorporate effects of recent major mergers \citep{BehrooziMajormergerVmax2014}.

In Table~\ref{tab:properties}, we list the properties of galaxies/halos 
defined in this subsection and the last.
Using these definitions, we obtain the characteristic times for each of the 
galaxies/halos in our sample, and we list them in 
Table~\ref{tab:sample}. In general, the piecewise functions 
introduced in \S\ref{ssec:galaxy-properties} fit well the evolution of the properties for most 
galaxies in our sample. 

There are three exceptional cases:
\begin{enumerate}[topsep=0pt,parsep=0pt,itemsep=0pt]
    \item {\objname m12z}: This galaxy starts to evolve toward colder dynamics,
    but the transition has not yet been completed by $z=0$. At $z=0$, its thin disk is still
    subdominant ($f_{\rm thin} \approx 0.25$), 
    and its hotness remains high ($\mathcal{H} \approx 0.5$). 
    The ending times, $t^{\text{(end)}}_{\text{thin}}$ and $t^{\text{(end)}}_{\mathcal{H}}$,
    are therefore undefined for this galaxy. 
    The star-formation burstiness of this galaxy never drops below $0.2$, 
    and its $t_{\mathcal{B}}$ is also undefined. These properties 
    are therefore not listed in Table~\ref{tab:sample}.
    \item {\objname m12w}: This galaxy has evolved to a dynamical state colder 
    than {\objname m12z}, with $f_{\rm thin} \approx 0.65$ and 
    $\mathcal{H} \approx 0.5$ at $z=0$. However, the transition appears 
    to still be ongoing, with the late-time plateau value ($f_2$ in 
    Eq.~\ref{eq:fitting}) for both $f_{\rm thin}$ and $\mathcal{H}$ 
    determined by only a few data points at 
    $t \gtrsim 12\Gyr$. The ending times, $t^{\text{(end)}}_{\text{thin}}$ and 
    $t^{\text{(end)}}_{\mathcal{H}}$, are therefore not well constrained, 
    and should be taken with caution for this galaxy.
    \item {\objname m12r}: The host halo of this galaxy has recent major 
    mergers at $z \lesssim 0.3$. This introduces significant late-time jumps 
    in the history of $V_{\rm vir}$ (and thus $\gamma$), causing the
    halo assembly to deviate from the typical two-phase pattern.
    This exception is also visible in the assembly-environment distribution 
    (Fig.~\ref{fig:Compare_largesim}), where the halo of {\objname m12r} 
    appears to be an outlier.
\end{enumerate}
Given the special status of these galaxies, we exclude them from the correlation analysis 
presented in \S\ref{sec:co-evolution}. The statistics shown there
should therefore only be interpreted as a result for the subset of galaxies whose halo assembly follows a typical two-phase 
pattern and whose galaxy dynamics has transitioned from hot to cold by $z=0$.
In \S\ref{sec:driving-processes}, we show that the transition of galaxy dynamics
depends on other factors in addition to halo assembly, and a more detailed discussion is 
presented on how these three galaxies can fit in the co-evolution scenario.

\section{Co-evolution pattern between galaxies and halos} 
\label{sec:co-evolution}

The time evolution of galaxy properties described above suggests a two-phase 
scenario for galaxy formation. This aligns well with the two-phase 
formation of dark matter halos in the $\Lambda$CDM cosmology 
\citep[e.g.][]{zhaoGrowthStructureDark2003,zhaoAccurateUniversalModels2009}. 
It is thus interesting to see whether the two phases of galaxies 
and halos are connected and, if so, how halo assembly affects the formation 
of galaxies in host halos. In this section, we examine the co-evolution 
between galaxies and halos phenomenologically, and in the next section,
we identify the physical processes that lead to the co-evolution. 
As we shall see, the connection between halo assembly and dynamical hotness of galaxies
leads to testable results for observations of galaxies in the early Universe.

\begin{figure*}
	\includegraphics[width=\textwidth]{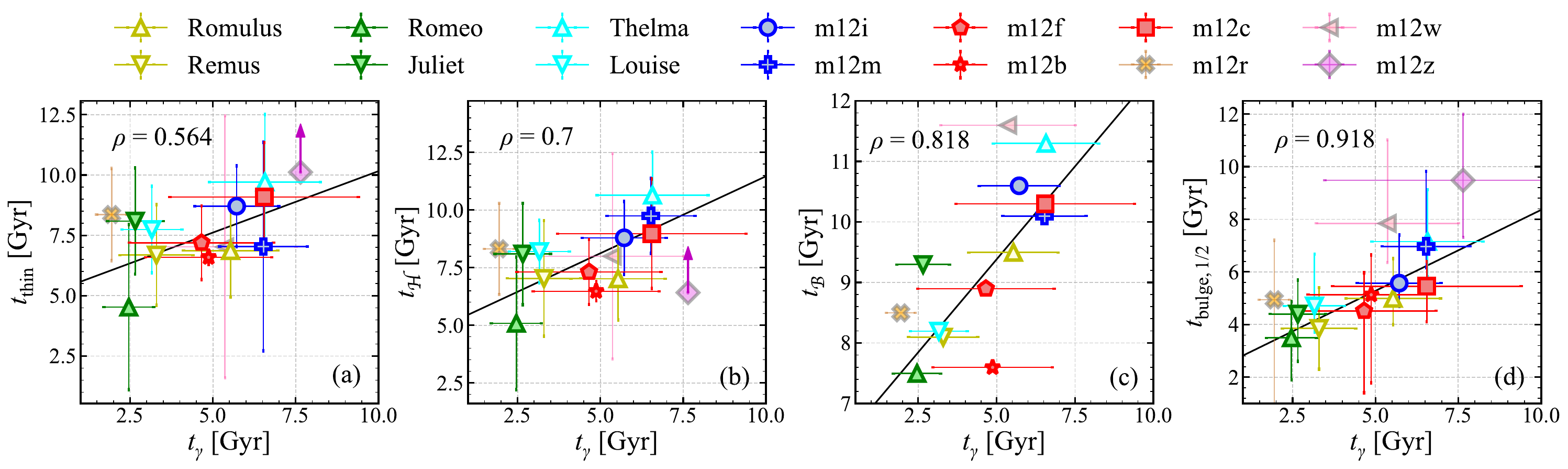}
    \caption{
        {\figem Correlation between halo transition time ($t_\gamma$) and galaxy 
        transition times.}
        Each panel uses one definition of galaxy transition time:
        {\figem (A)}, $t_{\rm thin}$;
        {\figem (B)}, $t_{\ssc \mathcal{H}}$; 
        {\figem (C)}, $t_{\ssc \mathcal{B}}$;
        {\figem (D)}, $t_{\rm bulge,1/2}$.
        Each marker represents one galaxy in our sample, with 
        error bar spanning from the start to the end of the transition interval
        (for $t_\gamma$, $t_{\rm thin}$ and $t_{\ssc \mathcal{H}}$) 
        or from  20\% to 80\% formation time of bulge mass (for $t_{\rm bulge,1/2}$).
        The ending epochs of the transitions of $f_{\rm thin}$ and 
        $\mathcal{H}$ are undefined for \texttt{m12z}, for which we 
        only show the starting epochs of the transitions
        with arrows pointing upward.
        For each pair of variables, the Spearman correlation coefficient ($\rho$)  
        is indicated in the corresponding panel (see also 
        Table~\ref{tab:corr} for a full list of $\rho$ and $p$-values),
        and black line shows a linear fitting.
        Markers with grey edges represent the exceptional galaxies
        (\texttt{m12w}, \texttt{m12r} and \texttt{m12z}) 
        excluded from correlation analysis and linear fitting.
        This figure demonstrates a co-evolution pattern between 
        halos and galaxies.
        See \S\ref{sec:co-evolution} for details.
    }
    \label{fig:Correlation}
\end{figure*}

\begin{figure}
	\includegraphics[width=0.48\textwidth]{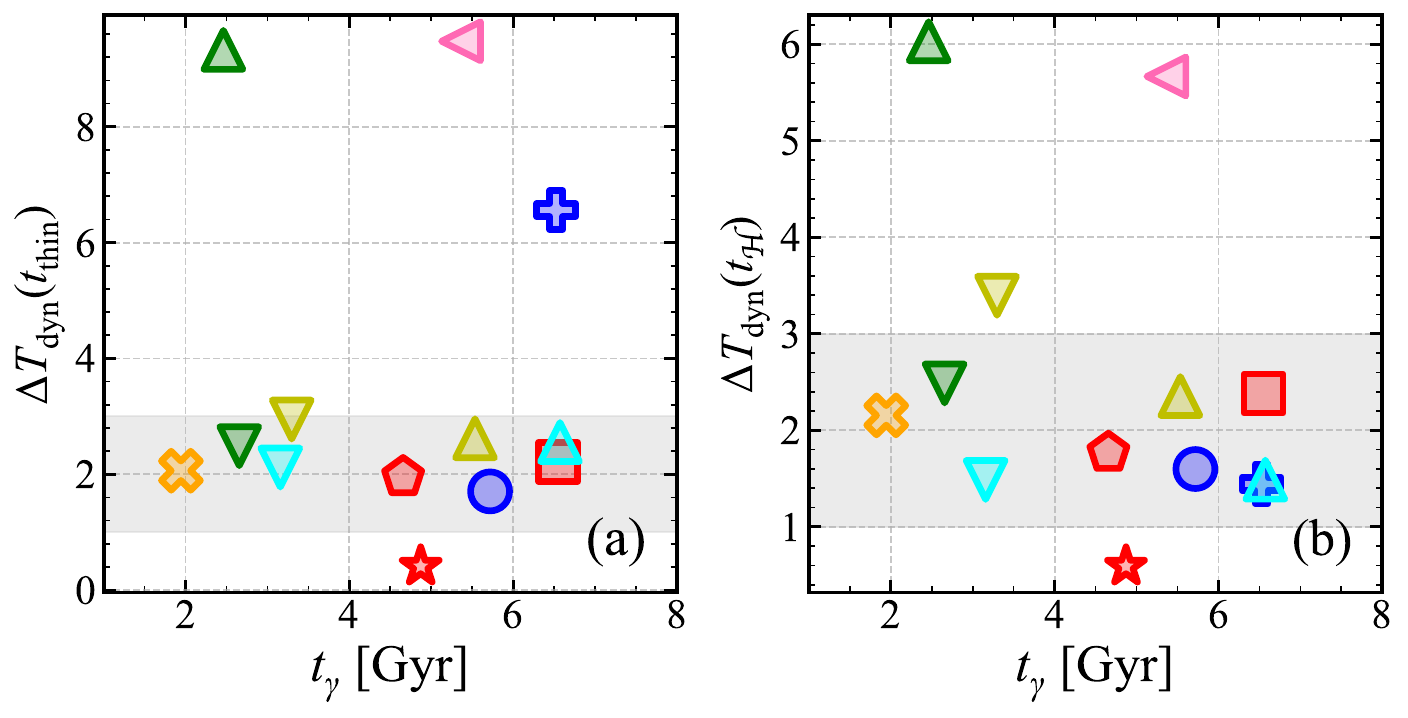}
    \caption{
        {\figem Durations of transitional phases for individual galaxies.}
        Here the the duration, $\Delta T_{\rm dyn}$, is normalized by the 
        dynamical timescale of the host halo (see Eq.~\ref{eq:duration}), and the transitional 
        phase is defined by the evolution of {\figem (a)}, $f_{\rm thin}$;
        {\figem (b)}, $\mathcal{H}$.
        Markers are the same as in Fig.~\ref{fig:Correlation}. 
        Grey shading indicates the range 
        $1 \leqslant \Delta T_{\rm dyn} \leqslant 3$.
        See \S\ref{sec:co-evolution} for details.
    }
    \label{fig:Dynamic_time}
\end{figure}

{\renewcommand{\arraystretch}{1.2}
\begin{table}
    \caption{{\figem 
    Spearman correlation coefficient ($\rho$) and $p$-value between 
    pairs of characteristic times of galaxies and their host halos.}}
    \label{tab:corr}
    \centering
    \tabcolsep=0.15cm
    \begin{tabular}{l|cccc}
        \hline
        $\rho$ ($p$-value) & $t_{\gamma}$ & $t_{\rm h, 1/2}$ & $t_{\mathrm{h, Vpeak}}$ & $t_{\mathcal{B}}$ \\
        \hline
        $t_{\mathrm{thin}}$ & 0.56 (0.07) & 0.30 (0.37) & 0.17 (0.61) & 0.85 (1e-3) \\
        $t{_\mathcal{H}}$ & 0.70 (0.02) & 0.27 (0.42) & 0.08 (0.81) & 0.85 (1e-3) \\
        $t_{\mathcal{B}}$  & 0.82 (2e-3) & 0.24 (0.48) & 0.30 (0.37) & -  \\
        $t_{\rm bulge,1/2}$  & 0.92 (7e-5) & 0.13 (0.69) & 0.53 (0.10) & 0.8 (3e-3)  \\
        \hline
    \end{tabular}

\end{table}
}

In Fig.\ref{fig:Correlation}(A)--(D), we show how the galaxy transition times 
($t_{\rm thin}$, $t_{\ssc \mathcal{H}}$, $t_{\ssc \mathcal{B}}$ and $t_{\rm bulge,1/2}$)
are correlated with the halo transition time ($t_\gamma$) using the 14 galaxies in our sample.
The solid line in each panel shows the linear fit to the points of individual galaxies, excluding 
the exceptional cases ({\objname m12w}, {\objname m12r} and {\objname m12z})
identified above. For every definition of the galaxy transition time, a positive correlation 
with $t_\gamma$ is observed: halos with earlier transition to slow accretion 
on average host galaxies with earlier transition to disk-dominated and dynamically cold
phases with lower level of burstiness in star formation. 
To quantify the significance of these correlations, we compute the 
Spearman correlation coefficient and the $p$-value between
each of the galaxy transition times and the halo transition time, 
and we list the results in Table~\ref{tab:corr}.
The correlation coefficient between $t_{\rm thin}$ and $t_\gamma$ is about 0.56,
with a $p$-value of 0.07, indicating a significant correlation.
The correlations of $t_{\ssc \mathcal{H}}$, $t_{\ssc \mathcal{B}}$ and $t_{\rm bulge,1/2}$ 
with $t_\gamma$ are even more significant and stronger, as reflected 
by the higher correlation coefficients of 0.70, 0.82 and 0.92, and 
the lower $p$-values of $0.02$, $2\times10^{-3}$, and $7\times 10^{-5}$, 
respectively. 
These results suggest that the transitions in galaxy morphology, dynamical 
hotness and star formation burstiness are closely related to the transition 
in halo assembly. A similar conclusion was reached by \citet{DillamoreARTEMISdiscform2024}
based on the correlation between the `spin-up' time, defined as the time when a galaxy 
starts to form ordered disk, and the halo assembly for MW-size galaxies in the ARTEMIS 
simulations.

In addition to the correlations between galaxy properties and halo assembly,
there are significant variations among individual galaxies, indicating that 
processes other than halo accretion may also play important roles in determining 
the transitions in galaxy properties. For example, the galaxy {\objname m12m} has 
a $t_{\rm thin}$ about $2\Gyr$ earlier than that expected from the linear fit. 
A careful examination reveals that {\objname m12m} has a massive satellite whose orbital 
motion aligns with the disk rotation of {\objname m12m} at $z \approx 3$. 
The transfer of orbital angular momentum, or the transfer 
of high-circularity stars from the satellite
to the disk may thus contribute to the early formation of the thin disk \citep{Jianghighzdiskgallaxy2025}. 
Similar variations were identified by \citet{chenHowEmpiricallyModel2021}
and \citet{chenMAHGICModelAdapter2021} by applying a set of successively refined regressors
to the galaxies in the IllustrisTNG and EAGLE simulations,
where nuanced factors other than halo assembly are inferred to affect 
star formation in galaxies.


For comparison, the correlations of the galaxy transition times with 
the half-mass formation time ($t_{\rm h,1/2}$) or the peak-velocity time
($t_{\rm h,Vpeak}$) of the host halo are much weaker, 
with $p$-values not low enough to reject the null hypothesis of no correlation. 
A similar result was reported by \citet{GarrisonFireIIgalkinematichalo2018}, 
who found a poor correlation between the stellar morphology of galaxies and 
the half-mass formation time of halos. These findings suggest that it is the 
transition in the halo accretion rate, rather than the accumulative amount 
of the accreted mass, that determines the transitions in galaxy properties.

Fig.~\ref{fig:Dynamic_time}(a) and (b) show the durations of the
transitional phases versus the halo transition time, $t_\gamma$,
for individual galaxies in our sample.
The duration of transition for each galaxy is normalized by the dynamical timescale of the 
host halo, $t_{\rm dyn, h}$, as 
\begin{equation}
    \Delta T_{\rm dyn} = \int_{t_1}^{t_2} \frac{\dd{t}}{t_{\rm dyn, h}(t)} 
    \,,\label{eq:duration}
\end{equation}
where the time interval 
$[t_1, t_2]$ is either $[t_{\rm thin}^{\rm (start)},t_{\rm thin}^{\rm (end)}]$
or $[t_{\ssc \mathcal{H}}^{\rm (start)},t_{\ssc \mathcal{H}}^{\rm (end)}]$.
The transitional phases often end within $1$--$3t_{\rm dyn, h}$, with a few number of 
exceptions that last longer. Typically, the timescales for the phase transition 
are much longer than the dynamic timescale of the galaxy itself (usually a few percent of 
$t_{\rm dyn,h}$), indicating that processes other than the dynamical adjustment of the 
galaxy are responsible for the completion of transitions, as we shall 
discuss in the following.


\begin{figure*}
    \includegraphics[width=\textwidth]{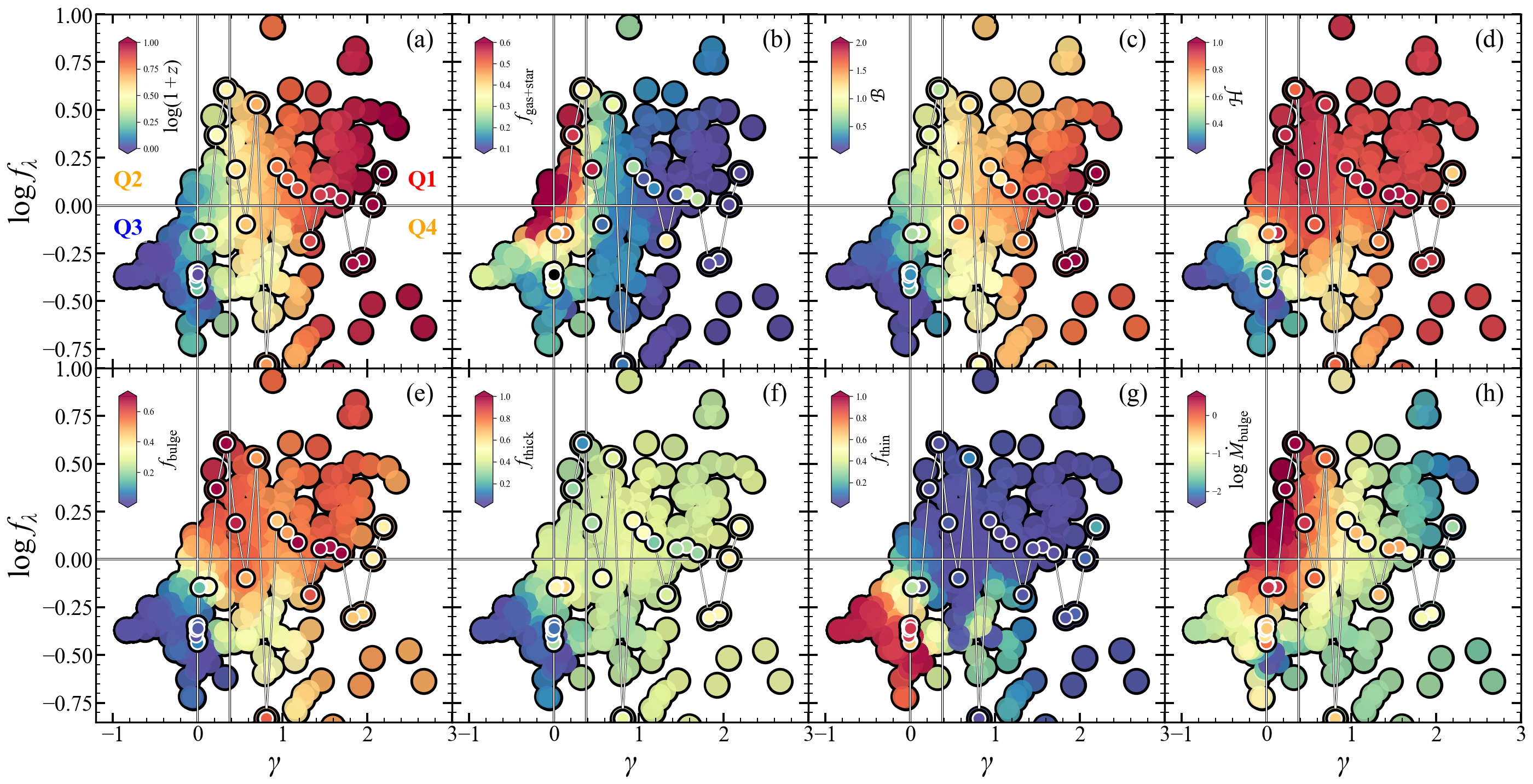}
    \caption{{\figem The quadrant diagram showing the evolution of galaxies 
    in the $f_{\lambda}$-$\gamma$ plane.}
    In each panel, dots show the snapshots of all galaxies in our sample,
    color-coded by a property labeled in the color bar.
    $f_\text{gas+star}$ in {\figem (b)} are evaluated within the inner 
    $1\Kpc$ of galaxies. The growth rate of stellar bulge 
    ($\dot{M}_{\rm bulge}$, in $\rm M_{\odot}yr^{-1}$)
    shown in {\figem (h)} is smoothed by the LOESS method \citep{Cleveland01091988}, using the Python package \citep{Cappellari2013}.
    White path linking the dots with black edges shows the 
    evolution of {\objname m12f}.
    Two vertical lines show $\gamma = 3/8$ and $\gamma = 0$, respectively,
    which define the transitional regime of halo assembly.
    One horizontal line shows $f_{\lambda} = 1$, above which self-gravity
    of gas is expected to prevent the formation of a dynamically cold disk.
    The $f_{\lambda}$-$\gamma$ plane is partitioned into 
    four quadrants, Q1--Q4, by these lines, within which galaxies are 
    expected to have distinct dynamical states. The FIRE-2 galaxies
    follow a general evolutionary trajectory in this plane, from 
    Q1 at high $z$ to Q3 at low $z$.
    See \S\ref{ssec:quadrant} for details.
    }
    \label{fig:quadrant}
\end{figure*}

\section{Physical processes driving the two-phase growth of galaxies}
\label{sec:driving-processes}

The correlation pattern between galaxies and halos found above 
motivates us to identify processes that drive the transition of 
galaxies based on the halo assembly. According to 
\citetalias{moTwophaseModelGalaxy2024} (see their \S3.3), 
two factors, one external and one internal, are predicted to be important in 
determining the dynamics of galaxies.

The first factor is the assembly rate of the host halo, as quantified by $\gamma$.
A halo in its fast phase has a nearly constant concentration of 
$c_{\rm v} \approx 4$ \citep{zhaoGrowthStructureDark2003,zhaoAccurateUniversalModels2009},
reflecting a self-similar growth without changing the shape of the profile.
Such an evolution of halo structure relies on a sufficient exchange of 
energy between the newly accreted matter and the existing matter in the halo. 
This is feasible only if the accretion is fast enough so that 
the perturbation of the gravitational potential is frequent and strong. 
The fast accretion of dark matter also has important implications for 
baryonic processes: fast accumulation of gas is expected to be associated with
fast accretion of dark matter. This, together with the effective radiative cooling 
expected before halos reach a threshold of $M_{\rm v} \approx 10^{12}\Msun$
(see the cooling diagram in, e.g., fig. 2 of \citetalias{moTwophaseModelGalaxy2024}), 
can cause large amounts of cold gas to fall almost freely into the halo center. 
Rapid inflow of gas sustains a high gas fraction that leads to the 
formation of a self-gravitating gas cloud (SGC) before angular-momentum 
support can be established.
Rapid change of gravitational potential is complemented by gas accretion and feedback 
processes in disturbing the gas and driving supersonic turbulence on 
galactic scale (as discussed in \S\ref{sec:introduction}).
Density fluctuations in the SGC are thus seeded and amplified, causing the formation of 
sub-clouds that are dense enough to be free of cloud-cloud collision and drag force 
from surrounding gas.
The almost ballistic motion of sub-clouds makes the mixing of angular momentum 
inefficient, prevents the formation of a dynamically cold disk and preserves the 
hot gas dynamics. Star clusters can form within the sub-clouds and 
inherit the hot dynamics, leading to the formation of a bulge-like stellar system. 
A dynamically cold gaseous and stellar disk is expected to form only after the halo assembly 
slows down ($\gamma \lesssim \gamma_{\rm f}$) and the gas fraction is reduced by, e.g. feedback, to a level that is 
comparable to the gas spin ($f_\lambda \equiv f_{\rm gas}/\lambda_{\rm gas} \lesssim 1$). This scenario predicts that the galaxy has hot 
(velocity-dispersion supported) dynamics at high $z$ and cold (rotation supported) 
dynamics  at low $z$.

The second factor is the relative importance between the self-gravity and 
angular momentum support of the gas, as quantified by $f_\lambda$, 
the ratio between the cold-gas fraction and the spin. This factor quantifies the 
internal instability of the disk. A high $f_\lambda$ can cause a global instability to 
develop in the disk, which is expected to thicken or destroy the cold disk even 
if it can be formed in the first place. Making $f_\lambda$ sufficiently small 
(again $f_\lambda \lesssim 1$)
is thus a necessary condition for the cold disk to form and to remain for a long
period.

\subsection{Evolutionary trajectory in the quadrant diagram}
\label{ssec:quadrant}

The combination of the two factors partitions the evolution of galaxies into 
four quadrants, Q1--Q4, according to their locations in the $f_\lambda$-$\gamma$ plane.
The dynamical states of galaxies in different quadrants are predicted to be distinct, 
as summarized below.
\begin{enumerate}[topsep=0pt,parsep=0pt,itemsep=0pt]
\item Q1: $\gamma \gtrsim \gamma_{\rm f}$ and $f_\lambda \gtrsim 1$.
This is where all galaxies were born at high $z$. A bulge-like morphology
is expected in this quadrant once stars form in the sub-clouds that move 
randomly. However, the concentration of the bulge is not necessarily high,
since stars may form in sub-clouds at large radii and dynamical heating 
associated with the fast halo assembly and star formation
can prevent the build-up of a concentrated structure.
\item Q2: $\gamma \lesssim \gamma_{\rm f}$ and $f_\lambda \gtrsim 1$.
This is where unstable disks are expected to form. External perturbations from 
the fast halo assembly are now alleviated, and mixing of angular momentum is 
elevated, both promoting the formation of a self-gravitating, disk-like structure.
However, local instability in the heavy disk can lead to the formation of 
clumps that can heat the disk dynamically \citep[e.g.][]{mengOriginGiantStellar2020,RomeoStellarmassfraction2020,
renaudGiantClumpsClouds2021,vandonkelaarGiantClumpsClouds2022,RomeoDiscgalaxiesangularmomentum2023}. 
Global instability of the disk can drive rapid gas inflow towards the galactic center to form a bar or pseudo-bulge with a certain level of dynamic hotness \citep[e.g.][]{duKinematicDecompositionIllustrisTNG2020,
zanaMorphologicalDecompositionTNG502022}.
%
\item Q3: $\gamma \lesssim \gamma_{\rm f}$ and $f_\lambda \lesssim 1$.
This is the regime in which a cold disk can form, since both external and internal 
barriers to the coldness are removed. To form a cold stellar disk, 
$f_{\rm gas}$ cannot be too low, because otherwise the gas cannot cool 
efficiently to form stars.
\item Q4: $\gamma \gtrsim \gamma_{\rm f}$ and $f_\lambda \lesssim 1$.
This is the regime where the gas fraction has been reduced by e.g. 
feedback processes, but dynamical heating by the fast halo assembly is still present.
Depending on the structure formed before this stage, the galaxy can be a bulge or 
thick disk.
\end{enumerate}

Note that the first factor, $\gamma$, is determined solely by the halo assembly
and is thus robust regardless of the detailed modeling of the baryonic physics in simulations.
However, the second factor, $f_\lambda$, depends strongly on baryonic physics, 
particularly on star formation, black-hole growth, and the feedback processes 
generated by them. Hydrodynamic simulations with distinct subgrid models can thus
predict diverse evolution pathways of galaxies in the $f_\lambda$-$\gamma$ plane. 
This also implies a rich set of possibilities for morphological transformation
when galaxies in the simulation migrate from one quadrant to another, 
thereby providing observational consequences to test models.

Fig.~\ref{fig:quadrant} shows the evolution of galaxies in the quadrant diagram.
In each panel, the dots represent individual galaxies at different $z$, 
color-coded by the quantity labeled in the color bar.
For reference, dots connected by lines in each panel of Fig.~\ref{fig:quadrant} 
show the evolution of {\objname m12f} in the quadrant diagram. 
As shown in the redshift evolution in Fig.~\ref{fig:quadrant}(a), 
the general evolutionary trend is that galaxies are born in Q1 
(large $\gamma$ and $f_\lambda$) -- the initial condition set by the $\Lambda$CDM 
cosmology, then repeatedly transit between Q1 and Q4 at around 
$f_\lambda \approx 1$, and eventually migrate into Q3.  
The repeated transitions between Q1 and Q4 are 
expected as a result of gas depletion and replenishment cycles associated 
with episodic feedback from the highly bursty star formation
\citep[e.g.][]{el-badryBreathingFIREHow2016,
wangVariabilityStarFormation2020,wangVariabilityStarFormation2020a,hopkinsWhatCausesFormation2023},
given the high level of burstiness in the early stage (Figs.~\ref{fig:Example}e,
and \ref{fig:Correlation}c; see also \citealt{muratovGustyGaseousFlows2015,
AnglesAlcazarCosmicBaryon2017,sternVirializationInnerCGM2021,
yuBurstyOriginMilky2021,yuBornThisWay2023}).
At low redshift ($z \lesssim 1$), $f_\lambda$ is reduced to below $1$ for most
galaxies owing to the reduction of gas supply in slow-accretion halos
and the continuous elevation of gas spin (Fig.~\ref{fig:Example}f; see also \S5.2
in \citealt{zjupaAngularMomentumProperties2017} for an explanation).
The fluctuation in $f_\lambda$ also becomes small, 
consistent with being influenced by the stabilized feedback strength from steady star formation 
and the deepened potential well that prevents the gas from being blown out
\citep[e.g.][]{hopkinsWhatCausesFormation2023}.
Fig.~\ref{fig:quadrant}(b) shows the evolution of $f_{\text{gas+star}}$,
a property that quantifies the self-gravity of a gas cloud. 
At both high and low redshifts, $f_\text{gas+star}$ is much smaller 
than 0.5, indicating that the gas is not globally self-gravitating.
The most interesting regime is the transitional phase ($\gamma \approx \gamma_{\rm f} = 0$--$3/8$), 
where $f_{\text{gas+star}}$ exhibits a significant increase and reaches a peak value of $\sim 0.5$,
indicating that the star-forming cloud in this regime becomes self-gravitating.
An examination of the evolution of $\dot{M}_{\rm bulge}$ 
(Fig.~\ref{fig:quadrant}h) shows that this regime coincides with
the interval where the formation rate of the bulge is the highest
among all redshifts. In fact, most bulge masses of galaxies in our sample are built 
up in this interval (see Fig.~\ref{fig:Example}a for example). 
The transitional phase represents a primary period for bulge assembly in 
the FIRE-2 galaxies,
as we will discuss in more detail in \S\ref{ssec:origin-hotness}. 
Note that fast formation of the bulge component also needs high 
$f_\lambda$ to ensure the condition of self-gravitating before 
angular-momentum support can be established, 
as seen in Fig.~\ref{fig:quadrant}(h).

Figs.~\ref{fig:quadrant}(c)--(g) show the evolution of the burstiness 
($\mathcal{B}$), hotness ($\mathcal{H}$), and the fractions of the 
three kinematic components (bulge, thick disk, and thin disk), respectively.
All these quantities exhibit obvious evolution when galaxies migrate from Q1/Q4 to Q3,
revealing a general trend that galaxies evolve from a dynamically hot phase,
with bulge-like morphology and highly bursty star formation, 
to a dynamically cold phase, with cold-disk morphology and steady star formation.
All these quantities show significant correlations with $\gamma$, 
and make transitions at $\gamma \approx \gamma_{\rm f} = 0$--$3/8$.
This confirms the findings in \S\ref{sec:co-evolution} that the two phases of 
galaxy evolution are correlated with the two phases of halo assembly.
A weak trend is evident that, given $\gamma$, galaxies with higher $f_\lambda$ 
tend to have a higher hotness and bulge fraction, verifying the expected 
importance of $f_\lambda$ in determining the dynamics of galaxies.
All these results support the use of the quadrant diagram to separate the 
phases of galaxy formation and to identify the physical processes 
underlying the transitions, 
as discussed in the following.

\begin{figure*} \centering
\includegraphics[width=0.8\textwidth]{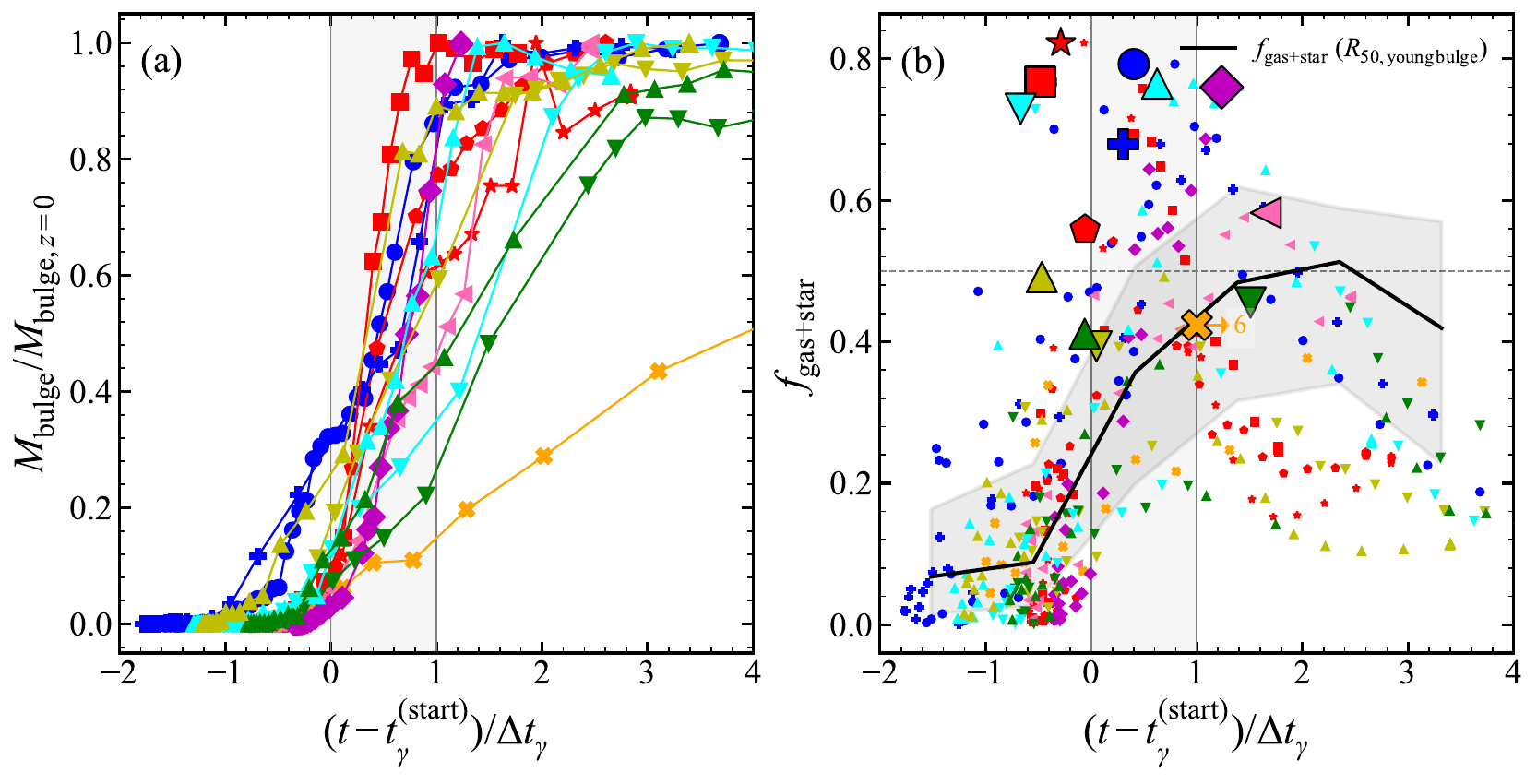}
    \caption{{\figem Evolution of properties related to the formation of stellar bulges.}
    {\figem (a)}, bulge mass ($M_{\rm bulge}$, normalized by the value at $z=0$) 
    as a function of time.
    {\figem (b)}, mass fraction of cold-gas plus young stars ($f_\text{gas+star}$) 
    within some apertures.
    Colored markers represent $f_\text{gas+star}$ evaluated within the 
    inner $1\Kpc$ in the history of one galaxy,
    as labeled in Fig.~\ref{fig:Correlation}.
    Black curve bound by grey shade shows the mean and 
    $25^{\rm th}$--$75^{\rm th}$ percentiles among all galaxies, 
    with $f_\text{gas+star}$ evaluated within the $R_{\rm 50,*}$ 
    of young bulge stars.
    The time variable of each galaxy is measured with respect to the 
    start time of transition of halo assembly, $t^{\text{(start)}}_{\gamma}$,
    and normalized by the duration of the transition, 
    $\Delta t_{\gamma} \equiv t^{\text{(end)}}_{\gamma} - t^{\text{(start)}}_{\gamma}$.
    We highlight the peak value of $f_{\text{gas+star}}$ along the 
    assembly history of each galaxy by a larger marker in {\figem (b)}.
    For {\objname m12r}, the peak of $f_{\text{gas+star}}$ is reached 
    at $t-t^{\text{(start)}}_{\gamma} \approx 6\Delta t_{\gamma}$, too 
    late to be shown, so we mark it with an orange arrow.
    Grey vertical band in each panel shows the transitional regime 
    of halo assembly.
    See \S\ref{ssec:origin-hotness} for details.
    }
\label{fig:Bulge_mass_tranistion}
\end{figure*}

\begin{figure*} \centering
\includegraphics[width=0.8\textwidth]{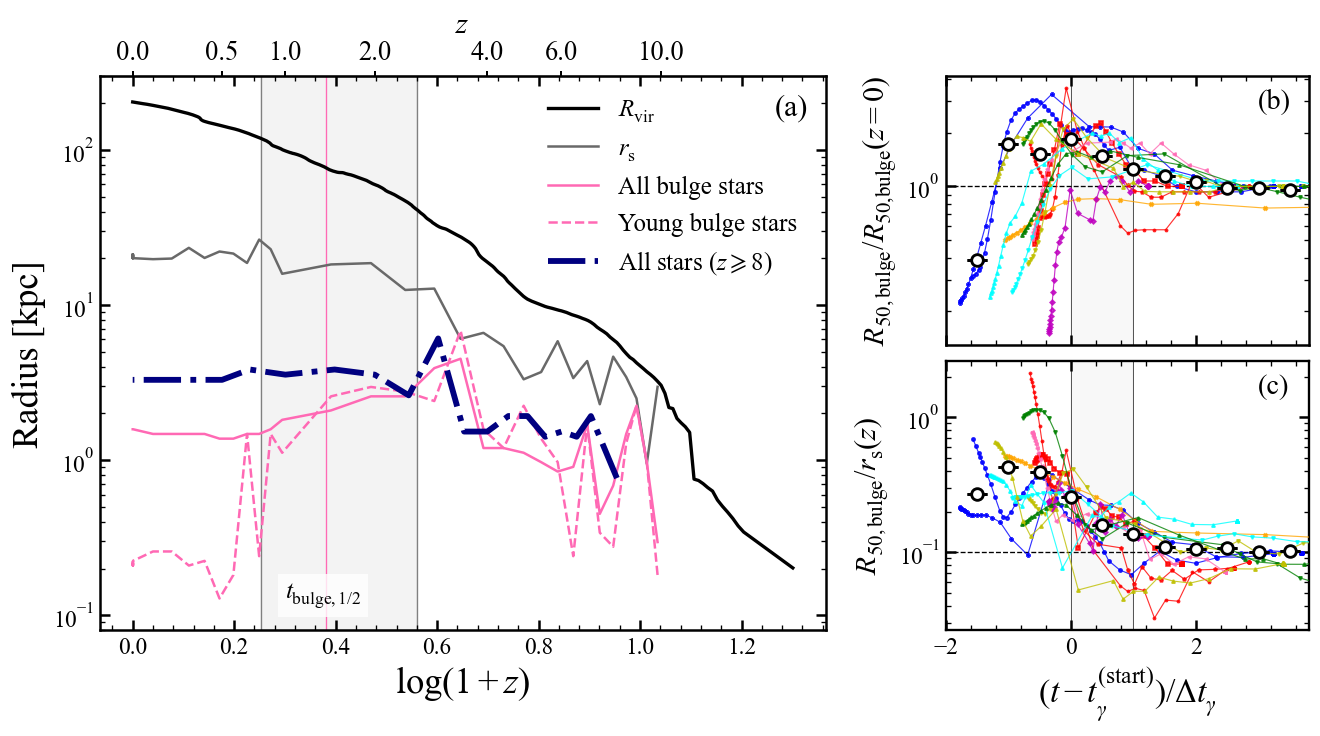}
\caption{{\figem Evolution of galaxy sizes.}
    {\figem (a)}, sizes of the galaxy {\objname m12f} by different definitions
    as functions of redshift,
    pink solid (dashed) for $R_{\rm 50,bulge}$ ($R_{\rm 50,young\,bulge}$), 
    the half-mass radius of all (young) bulge stars,
    and {\figem dark blue} for the half-mass radius of all stars formed 
    at $z \geqslant 8$.
    For comparison, we show the evolution of virial radius,
    $R_{\rm vir}$ (black), and scale radius 
    $r_{\rm s} \equiv R_{\rm vir}/c$ (grey), of the host halo,
    where $c$ is the concentration parameter obtained by fitting 
    the radial density profile with the NFW form \citep{navarroStructureColdDark1996,navarroUniversalDensityProfile1997,bullockProfilesDarkHaloes2001}.
    Pink vertical line indicates $t_{\rm bulge,1/2}$, 
    the half-mass formation time of the stellar bulge.
    {\figem (b)}, {\figem (c)},
    $R_{\rm 50,bulge}$ of all galaxies in our sample,
    normalized by either the value at $z=0$ ({\figem b})
    or the scale radius of their host halo at the corresponding 
    time ({\figem c}).
    The time variable of each galaxy is 
    measured with respect to $t^{\text{(start)}}_{\gamma}$
    and normalized by $\Delta t_{\gamma} \equiv t^{\text{(end)}}_{\gamma} - t^{\text{(start)}}_{\gamma}$.
    Black circles show the mean values among all galaxies.
    The markers for each galaxy are styled the same as Fig.~\ref{fig:Correlation}. 
    In each panel, grey vertical band shows the transitional regime 
    of halo assembly.
    See \S\ref{ssec:origin-hotness} for details.
    }
\label{fig:radius}
\end{figure*}

\subsection{The origin of dynamical hotness: concentrated versus 
scattered star formation}
\label{ssec:origin-hotness}

The above analyses based on the quadrant diagram pinpoint two regimes for the
build-up of the dynamically hot stellar system in a galaxy. 
The two regimes are well separated
in the $f_\lambda$-$\gamma$ plane, and exhibit distinct patterns in the physical 
properties of galaxies. 
The later regime is at $\gamma \approx 0$--$3/8$, where the fraction of 
cold-disk stars is about to rise (see e.g. Fig.~\ref{fig:Example}a),
but the SFR of the bulge component is the highest
among all redshifts in the history of a galaxy (see Fig.~\ref{fig:quadrant}h).
In Fig.~\ref{fig:Bulge_mass_tranistion}(a), we show the assembly histories 
of the bulge mass for all 14 galaxies in our sample.
All of these assembly histories have a sigmoid-like shape, 
with clear transition during the transition phase of halo assembly,
$t_\gamma^{\rm (start)} \lesssim t \lesssim t_\gamma^{\rm (end)}$.
Bulge masses grow rapidly in this duration, and most bulge stars 
at $z = 0$ form during this period.
The large correlation coefficient, $\rho = 0.918$,
between bulge formation time ($t_{\rm bulge,1/2}$) 
and halo transition time ($t_\gamma$) shown in Fig.~\ref{fig:Correlation}(D)
is thus explained.
The earlier regime is at $\gamma \gtrsim 3/8$, before the halo 
assembly slows down. Star formation
in this regime contributes only a minor fraction of the final bulge mass
at $z = 0$ (see Fig.~\ref{fig:Bulge_mass_tranistion}a), 
but almost all stars are formed in dynamically hot components 
(bulge or thick disk; see Figs.~\ref{fig:Example}c and \ref{fig:quadrant}e, f).
The two regimes thus have important implications both for 
observations targeting the old bulge stars at low $z$ and for deep
observations targeting the ongoing bulge formation at high $z$, 
as we will discuss in \S\ref{sec:early-growth}.

The evolutionary pattern of galaxies in the quadrant diagram also provides 
clues to identify the origin of the hotness in each of the two regimes.
One critical feature for the properties of a galaxy in the later regime of bulge formation
is that the star-forming gas in the inner region 
of the galaxy is globally self-gravitating.
This is evident from Fig.~\ref{fig:quadrant}(b), where 
the fractions of cold gas plus young stars
($f_\text{gas+star}$) in the inner $1\Kpc$ of the galaxies 
are seen to reach $0.5$ when $\gamma \approx 0$--$3/8$,  
indicating a self-gravitation of the star-forming gas.
To see this more clearly for individual galaxies, in 
Fig.~\ref{fig:Bulge_mass_tranistion}(b), we show
the evolution of $f_{\text{gas+star}}$ for all the 14 galaxies 
in our sample, with the time axis aligned to $t_\gamma^{\rm (start)}$ 
and scaled by the duration of the transitional phase,
$\Delta t_\gamma \equiv t_\gamma^{\rm (end)} - t_\gamma^{\rm (start)}$.
The radius to measure $f_{\text{gas+star}}$ is chosen to be either 
$R_{\rm 50,young\,bulge}$, the half-stellar-mass radius of young bulge stars
(black curve with gray shade) or $1\Kpc$ (scatter points).
Before $t_\gamma^{\rm (start)}$, $f_{\text{gas+star}}$ is 
low in most cases, indicating that no global self-gravitation of 
the star-forming gas is present. Between
$t_\gamma^{\rm (start)}$ and $t_\gamma^{\rm (end)}$, 
$f_{\text{gas+star}}$ increases significantly and reaches about $0.5$,
indicating that the star-forming gas becomes globally self-gravitating.
At $t \gtrsim t_\gamma^{\rm (end)}$, 
$f_{\text{gas+star}}$ within $1\Kpc$ drops again to a low value,
while $f_{\text{gas+star}}$ within $R_{\rm 50,young\,bulge}$ 
remains close to $0.5$. This suggests that the star-forming gas
responsible for the bulge formation is still globally self-gravitating,
but its size has shrunk below $1\Kpc$, a value much smaller 
than the size of the thin disk component that dominates 
the star formation at low $z$ (see also Fig.~\ref{fig:radius} to be
discussed below). Bulge formation in this later stage, $t \gtrsim t_\gamma^{\rm (end)}$, 
is highly inefficient
and contributes only a small fraction to the instantaneous SFR and the final 
bulge mass.

The above analyses suggest the origin of hotness in the later regime of bulge formation. 
Once the spatial distribution of the star-forming gas in a galaxy becomes 
concentrated enough so that the self-gravitation condition is satisfied,
violent fragmentation of the gas naturally arises owing to 
various types of instabilities, such as the Jeans instability. 
Sub-clouds formed in this way then produce star clusters, which 
inherit the hot dynamics of the gas originated from 
the perturbations and turbulence
associated with the build-up of the high gas concentration.
The total mass formed
in this way can be very large as a result of the rich amount of 
gas accumulated during the fast phase of halo assembly and 
the deepened gravitational potential well of the host halo
that may resist gas loss by feedback. The formation of the stellar bulge 
proceeds so rapidly that a dynamically cold gaseous disk does not have 
enough time to form. 
Given the importance of high gas concentration in driving the formation of 
stellar bulges in this regime,
we summarize the corresponding origin of the hotness
as `{\bf\em concentrated star formation}'.
In Fig.~\ref{fig:radius}(a), we show the evolution of the half-stellar-mass 
radii of different stellar components of {\objname m12f}, 
in comparison to the scale radius ($r_{\rm s}$) and virial radius ($R_{\rm vir}$) 
of the host halo. 
Between $t_\gamma^{\rm (start)}$ and $t_\gamma^{\rm (end)}$,
the growth of $R_{\rm vir}$ continues, but the 
size of the distribution of young bulge stars decreases significantly. 
The small size of newly formed bulge stars in 
comparison with the large halo size confirms the scenario
of concentrated star formation inferred above from self-gravitation.
At $t \gtrsim t_\gamma^{\rm (end)}$,
the size of newly formed bulge stars drops to a few 
hundred parsecs, indicating that the concentrated star formation
continues to take place, albeit at a very low level and within 
a very small region.
The behavior in the evolution of the bulge size seen in {\objname m12f},
is also shared by other galaxies, as shown by Fig.~\ref{fig:radius}(b).
In all cases, the bulge radius first increases rapidly, reaching 
a maximum value at $t \approx t_{\gamma}^{({\rm start})}$,
and then decreases slowly with time, eventually reaching a roughly constant 
value at $t> t_{\gamma}^{({\rm end})}$. As shown in Fig.~\ref{fig:radius}(c), when normalized 
by the scale radius of the host halo at the redshift when $R_{\rm 50,bulge}$ is 
measured, the ratio is roughly a constant of $\sim 0.3$ at $t<t_{\gamma}^{\rm (start)}$, 
with large fluctuations generated by the rapid change of the gravitational potential.
The ratio then decreases by a factor of about three and settles around $\sim 0.1$ 
at $t>t_{\gamma}^{\rm (end)}$. As described in \citet{chenTwophaseModelGalaxy2024},
in the two-phase model of galaxy formation, the bulge size is related 
to $r_{\rm s}$ as $R_{\rm 50,bulge} \approx \alpha r_{\rm s}$, with $\alpha$
being a redshift- and mass-independent constant that is determined by the 
self-gravitation condition and bulge profile.
Matching the prediction of the model with the observational data, \citet{chenTwophaseModelGalaxy2024}
obtained $\alpha\approx 0.04$. As one can see from Fig.~\ref{fig:radius}(c), 
most of the simulated galaxies follow the model scaling relation after the 
concentrated formation of the bulge, but their values of $\alpha$ are larger 
by a factor of more than two. This discrepancy may partly arise from the fact 
that simulated bulges include all stars in the outer region where the surface 
brightness may be too low to be account for in the observational data. 
Another possibility is that the bulge to halo mass radios in simulated galaxies 
are larger than in real galaxies. Indeed, the stellar to halo mass ratios 
of FIRE-2 or FIRE-set galaxies (see Table~\ref{tab:sample}; 
see also \citealt{hopkinsFIRE2SimulationsPhysics2018,SchayeCOLIBRE2025})  are more than a factor of two 
higher than in observations \citep[e.g.][]{yangEvolutionGalaxyDarkMatter2012}. 

In the earlier regime of bulge growth, the star-forming gas is not 
globally self-gravitating, as seen from the low $f_{\text{gas+star}}$ 
in Fig.~\ref{fig:Bulge_mass_tranistion}(b) during
$t \lesssim t_\gamma^{\rm (start)}$.
This, in combination with the high gas fractions in, 
and the small sizes of halos,
suggests that star-forming gas must be distributed in an extended way
and that the bulge size-halo size ratio is much higher than
that in the later regime of bulge formation.
The evolution of sizes in Fig.~\ref{fig:radius}(a) shows 
this point, where $R_{\rm 50,young\,bulge}/R_{\rm vir}$ 
remains as high as $0.1$ -- a value much higher than 
$0.01$--$0.015$ observed for bulges 
in the local Universe \citep[e.g.][]{kravtsovSizeVirialRadiusRelation2013,
huangRelationsSizesGalaxies2017,zanisiGalaxySizesGalaxyhalo2020,
chenTwophaseModelGalaxy2024}.
These results suggest that the origin of hotness in the earlier regime is 
that the distribution of star-forming gas is sufficiently extended so that 
the random motion of gas has not been significantly dissipated until 
stars are born out of the gas. 
This is totally different from the situation where the angular momentum 
between gas elements is well 
mixed to produce a dynamically cold gaseous disk with ordered rotation
before most of that gas is converted into stars.
This is, however, similar to the situation in the formation of 
the brightest cluster galaxies (BCGs) in massive halos at low $z$,
where significant fractions of stars were formed ex-situ in 
satellite galaxies and then accreted to the central galaxies
to build their extended stellar outskirts \citep[e.g.][]{bezansonRelationCompactQuiescent2009,
HuangHSCstellarhaloes2018,klugeMinorMergersAre2023}.
Given the large extension of the star-forming gas and its importance
in determining the hot stellar dynamics, we refer to the origin of 
the hotness in this regime as `{\bf\em scattered star formation}'. 

\begin{figure*}
	\includegraphics[width=0.92\textwidth]{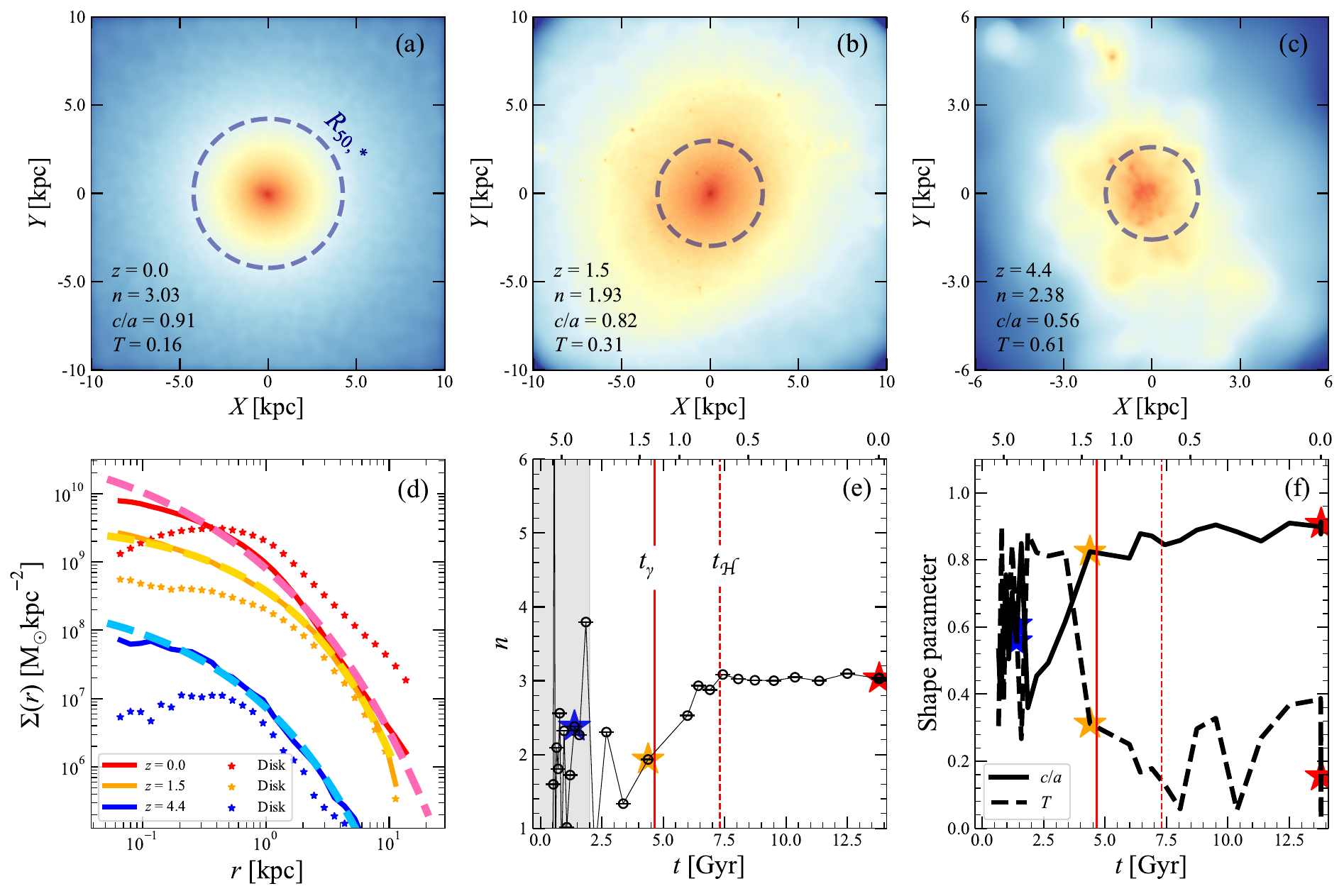}
    \caption{{\figem Evolution of the structure of the stellar bulge of \texttt{m12f}.}
    {\figem (a)}--{\figem (c)}, 2-D stellar mass surface density
    of the bulge component at three redshifts, respectively.
    Dashed circle in each panel indicates $R_{\rm 50,bulge}$.
    {\figem (d)}, stellar mass surface density as a function of galactocentric
    distance ($r$), evaluated using bulge stars from the simulation (solid) 
    and fitted with a S\'ersic profile (dashed) at three redshifts.
    For comparison, starred markers show the profile using disk (thin + thick) stars.
    {\figem (e)}, S\'ersic index of the stellar bulge as a function of cosmic time.
    {\figem (f)}, minor-to-major axis ratio ($c/a$; solid) 
    and triaxiality ($T$; dashed) of the stellar bulge as functions of cosmic time.
    In {\figem (e)} and {\figem (f)}, vertical red solid (dashed) line represents 
    $t_{\gamma}$ ($t_{\mathcal{H}}$).
    Starred markers indicate the three redshifts displayed in {\figem (a)}--{\figem (d)},
    and the corresponding structural parameters are labeled
    in {\figem (a)}--{\figem (c)}.
    In {\figem (e)},
    gray vertical band marks the range of redshift where the S\'ersic fitting 
    is ill-defined due to the temporal and spatial
    fluctuations in the surface density profiles.
    This figure highlights that bulge evolves from a prolate, clumpy and
    low-concentration structure at high $z$ to a spherical, smooth and
    high-concentration structure at low $z$. See \S\ref{sec:early-growth} 
    for details. 
    }
    \label{fig:bulge_illustration}
\end{figure*}

As suggested by numerical simulations 
\citep[e.g.][]{keresHowGalaxiesGet2005,Dekelcoolflowshockheating2006,dekelColdStreamsEarly2009,danovichFourPhasesAngularmomentum2015}, 
galaxies at high $z$ are often fed by cold streams of gas
that are seeded by large-scale dark-matter filaments
and condensed by radiative cooling. Scattered star formation is thus 
expected to take place in these cold streams. 
This is indeed the case for FIRE-2 galaxies, as demonstrated by the 
example shown in Fig.~\ref{fig:bulge_illustration}(c).
The environment provided by a fast-accreting halo at high $z$
can thus be understood as to provide a condition for the 
IGM/CGM similar to that satisfied in satellite galaxies in massive halos at 
low $z$, where gas density may be 
sufficiently high to allow local self-gravitation 
and star formation. Note that the large extents and filamentary structures 
of newly formed stars also predict features in the size and morphology of high-$z$ 
galaxies directly testable by deep observations,
as we will discuss in \S\ref{sec:early-growth}.

Given the two distinct modes of star formation that build up the dynamically hot
stellar component, a question arises as what drives the transition of galaxies from 
the scattered star formation to the concentrated star formation. This again can 
be inferred from the evolution of galaxies in the quadrant diagram.
An obvious difference between the two regimes is the assembly rate 
of the host halo. As suggested by N-body simulations \citep{BorzyszkowskiZOMG2017}, 
fast-accreting halos are often connected with filaments
in the cosmic web, while slow-accreting halos are not.
The cold streams feeding the formation of bulges in the earlier regime 
($\gamma \gtrsim 3/8$) are thus expected to disappear in the later 
regime ($\gamma \approx 0$--$3/8$), 
as dark matter filaments seeding the gaseous streams disappear.
Consequently, the high density condition for scattered star formation
in CGM/IGM may not be satisfied
in the later regime, allowing the gas to accumulate into 
the halo center without being consumed by star formation and feedback
until the condition of self-gravitation is satisfied globally to trigger an 
intense and concentrated star formation.
During the transitional phase, the potential well of the host halo,
as quantified by the maximum circular velocity, $V_{\rm max}$, 
has already reached the peak value among all redshifts 
\citep{zhaoGrowthStructureDark2003}. This, together with
the rich amount of gas accumulated during the fast phase and 
the self-gravity of the gas that further deepens the potential well,
creates a condition that is highly resistant to feedback 
\citep{hopkinsWhatCausesFormation2023}.
A large amount of gas can thus be converted into stars in a short period,
which explains the rapid growth of bulge mass during the concentrated 
star formation.

Another mechanism that may also contribute to the transition from 
the scattered to the concentrated star formation is `dynamical heating',
by which stars formed at early epochs are `heated' later on
by gravitational perturbations.
One source of dynamical heating is the fast assembly of the dark matter
halo, whose effect is expected to be strong at high $z$. 
The large extents of bulges formed via scattered star formation at high $z$ 
imply that the total gravity on the bulge is dominated by that of the  
dark matter. As halos in the fast phase self-similarly expand their 
structure \citep{wechslerConcentrationsDarkHalos2002,zhaoGrowthStructureDark2003}, 
stars co-expand with dark matter
as they gain energy from the fast-varying gravitational potential.  
In Fig.~\ref{fig:radius}(a), we show the evolution of the half-mass radius 
of the stars formed at $z \geqslant 8$ (dark-blue curve). 
The half-mass radius of these old stars expands with time,
suggesting that dynamical heating is indeed effective.
The long-term trend of such an expansion follows that of the halo
radius, indicating that dynamical heating by fast halo assembly
is likely an important heating mechanism.
The other sources of dynamical heating can be from
episodic outflow produced by bursty star formation 
that causes temporary loss of binding energy of stars, 
and the formation of dense gas/stellar clumps that scatter stars
and transfer energy to them. Such effects have been suggested by e.g.
\citet{el-badryBreathingFIREHow2016}, who 
showed that bursty star formation leads to outward migrations of stars, 
more so for older stars \citep{KadoFongFIREdwarfstellaroutskirts2022}, 
and by \citet{mccluskeyDiscSettlingDynamical2024}, 
who demonstrated that the velocity dispersion of stars
rapidly increases with time at high $z$.
Separating the sources of dynamical heating, however, is 
not straightforward, as both dark-matter and baryonic 
sources of dynamical heating are expected to be more 
effective at higher $z$. Future studies utilizing idealized simulations
can help to disentangle these sources, and to quantify their relative 
importance. 

While the transition between the two regimes of bulge formation
is clearly seen in our sample of galaxies, we note that this
largely relies on the specific sub-grid models adopted in the FIRE-2 simulations: 
once gas is accreted into a halo through filaments and condensed
to a density sufficient for star formation, the `early' feedback produced
by young stars through, e.g. stellar winds and ionizing radiation, may be
able to disperse the gas before more stars can form.
Such feedback from star formation has been investigated in some 
hydrodynamic simulations for isolated galaxies
\citep{Ublerfeedback2014,AgertzFeedbackonSFR2015,AgertzFeedbackongalaxy2016,
liEffectsSubgridModels2020,dengRIGELSimulatingDwarf2024}, 
and is expected to be important also in cold filaments due to their
shallower gravitational potential well than that in the ISM.
In FIRE-2, the early feedback was included in the form of internal radiative feedback from the stars and external radiative feedback from the UV background, 
but the implemented feedback strength may not be sufficient to prevent 
star formation in the filaments. Future observations able to
detect the low-surface-brightness outskirts of high-$z$ galaxies
are crucial to test the presence of the scattered star formation 
and to narrow down sub-grid models to be implemented in simulations.

While fast halo accretion is related to bulge formation in our sample of
FIRE-2 galaxies, it should not be considered as a necessary condition for
galaxies to be dynamically hot. Other ``heating'' sources may also disturb galaxies 
and contribute to the dynamical hotness. For example, galaxy-galaxy mergers and close encounters 
were found to be critical in shaping the structure of massive
elliptical galaxies at late times \citep[e.g.][]{shenSizeDistributionGalaxies2003,
NaabMinorMergersMassive2009,klugeMinorMergersAre2023}; 
the changes in the
gravitational potential caused by inflow/outflow in intense star-formation
cycles were proposed as a mechanism to produce the structural diversity of
dwarf galaxies \citep[e.g.][]{el-badryBreathingFIREHow2016,dicintioNIHAOXIFormation2017}; 
rapid and/or filamentary gas accretion can provide additional dynamical heating beside the
gravity of dark matter \citep{Dekelcoolflowshockheating2006,RobertsonAdiabaticHeatingContracting2012,ginzburgEvolutionTurbulentGalactic2022,GoldnerAccretionDrivenTurbulence2025};
more exotic heating mechanisms, such as dark-matter self-interaction
\citep{BalbergSelfInteractingDark2002,zhangUnexpectedClusteringPattern2025}, may also produce dynamical hotness. 
The fast halo accretion is neither a sufficient condition for
galaxies to be dynamically hot. As suggested by
\citet{moTwophaseModelGalaxy2024}, one additional condition for the formation of dynamically hot 
galaxies is that gas cooling in the CGM is rapid so as to sustain gravitational instability 
and active star formation. In Appendix~\ref{app:CGM-virialization},
we show that the cooling time is indeed shorter than the free-fall time during
the fast-accretion phase for galaxies in our sample, so that the condition of 
rapid cooling is satisfied.



\subsection{Processes driving the transition to cold disk}
\label{ssec:process-transition}

As discussed above in connection to the quadrant diagram, the formation of a 
thin disk requires two conditions. First, the gravitational potential well
should be stable so that dynamical heating by variation of the gravitational
potential is not important. 
A high accretion rate of the halo, represented by a high value of $\gamma$, 
can perturb the gravitational potential and dynamically heat baryons
on galactic scales. 
Thus, the formation of thin disks
occurs predominantly when $\gamma$ is low, consistent with the correlation between $t_{\rm thin}$ 
and $t_\gamma$ shown in Fig.~\ref{fig:Correlation}. 
The second condition is that the cold gas fraction must be sufficiently low 
so that the total gravitational potential in the inner region of the halo 
is not dominated by the disk. This is demonstrated by the correlation between 
$f_{\rm thin}$ and $f_\lambda$ shown in Fig.~\ref{fig:quadrant}(g), and by
the rapid decline of $f_{\text{gas+star}}$ within inner $1\Kpc$ 
at $t>t_\gamma^{({\rm end})}$,  
shown in Fig.~\ref{fig:Bulge_mass_tranistion}(b), when the star formation starts 
to be dominated by that in the thin disk (see  Fig.~\ref{fig:Example}). 

In the simulated galaxies, the gravitationally dominant component undergoes two transitions. 
Initially, at $t< t_{\gamma}^{({\rm start})}$ when $\gamma$ is large, 
dark matter dominates the gravitational potential almost everywhere
in the halo, as implied by the large sizes of galaxies with respect to 
the sizes of halos (see Fig.~\ref{fig:radius}c). During this phase, 
a small amount of stars is formed in the `scattered mode' discussed 
in \S\ref{ssec:origin-hotness}, and the gravitational potential near the 
halo center is still dominated by dark matter. However, as the evolution 
moves to the transitional phase of halo assembly, $t_{\gamma}^{({\rm start})}<t< t_{\gamma}^{({\rm end})}$, 
a large amount of low-spin gas accumulates in the halo center, becomes self-gravitating, 
forms bulge stars in the `concentrated mode', and reduces the gas by feedback.
As shown in Fig.~\ref{fig:Bulge_mass_tranistion}(b), 
in almost all simulated galaxies, the gravitational potential in the central 
part of the halo becomes dominated by star-forming gas and young stars 
in the bulge in this phase. 
As soon as the bulge component is established at $t>t_{\rm bulge,1/2}\sim t_{\gamma}$, 
it provides another source to stabilize the gravitational potential and to
promote the formation of a dynamically cold disk from gas with
sufficiently high spin \citep[see also][]{hopkinsWhatCausesFormation2023}. 
As the disk grows, the bulge component can continue to grow slowly from
low-spin gas near the center, as shown in Fig.~\ref{fig:visualization}f.
Thus, the transition to thin-disk formation in the simulated galaxies is well 
described by the moving of a galaxy to Q3 in the quadrant diagram. 
In particular, the formation of a dynamically hot bulge in the fast accretion
regime of the halo promotes the formation of a thin disk by stabilizing the
gravitational potential as well as by reducing the amount of low-spin gas, 
making the transition to thin-disk formation at $t> t_{\gamma}^{({\rm end})}$
more prominent.

As described in the end of \S\ref{sec:sim}, the three galaxies, {\objname m12w}, 
{\objname m12z} and {\objname m12r}, are exceptional in their transitions in galaxy dynamics and halo assembly. 
These exceptions point to the presence of additional heating sources, such as those 
discussed at the end of \S\ref{ssec:origin-hotness}, that can delay 
the dynamical settling and therefore deserve special attention. 
Galaxies {\objname m12w} and {\objname m12z} maintain a high gas
fraction after the transitional phase of halo assembly (see the pink triangles
and purple squares in Fig.~\ref{fig:Bulge_mass_tranistion}b). 
A more detailed examination shows that their values of $f_\lambda$ are close to 
$1$ at $z \approx 0$. These two galaxies thus never entered Q3 of the 
quadrant diagram, implying that the condition of gas self-gravitating
is not removed and the growth of a dynamically cold disk is prohibited.
{\objname m12r} is similar to these two galaxies in that it also maintains 
a high gas fraction after the transitional phase of halo assembly.
The difference is that the halo of {\objname m12r} made a transition to the slow 
phase very early and has already entered Q3 at $z \approx 0$.
The disk of {\objname m12r} is therefore capable of growing and dominating 
the stellar mass at $z = 0$.
Interestingly, the host halo of {\objname m12r} has a recent major merger (with a 
halo mass ratio of about $1:3$) at $z \approx 0.3$, which brings into the system 
a satellite galaxy that merges with {\objname m12r} at $z \approx 0.22$. 
This satellite has an orbit almost co-rotating with the disk of {\objname m12r}, 
which allows the disk of {\objname m12r} to survive the merger.
At $z \approx 0.1$, another halo merges into the host halo of {\objname m12r}
(with a halo mass ratio of about $1:7$). This introduces a close companion 
to {\objname m12r} at $z \approx 0$ and a galaxy-galaxy merger may occur in the near future.
A question remains as to how these three exceptions maintain a high gas fraction 
after the transition of halo assembly. The distribution of these galaxies in 
the assembly-environment planes (Fig.~\ref{fig:Compare_largesim}) shows that they are  
among the galaxies with the lowest environmental density and that {\objname m12r} has 
an exceptionally late $t_{\rm h,1/2}$. Furthermore, the halo masses of these galaxies are also 
relatively low (see Table \ref{tab:sample}). These properties may result in different gas inflow 
and outflow in FIRE-2, so that a high gas fraction can be sustained even at late 
times.

\begin{figure*}
	\includegraphics[width=0.97\textwidth]{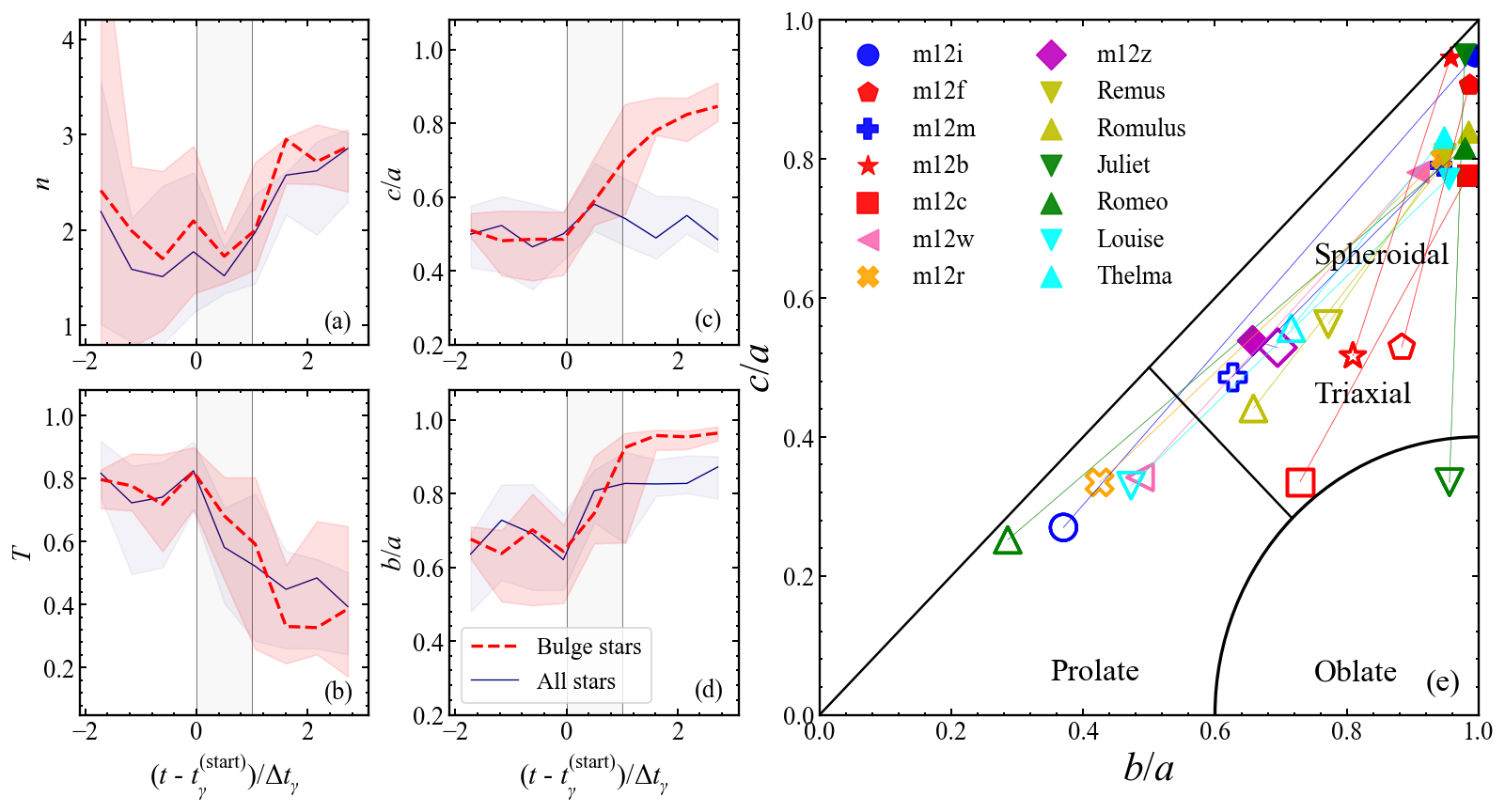}
    \caption{{\figem Evolution of structural parameters of stellar bulges.}
    {\figem (a)}--{\figem (d)}, evolution of
    the S\'ersic index ($n$), triaxiality ($T$), and axis ratios $c/a$ and 
    $b/a$, respectively, of the stellar bulges
    (red; curve for the mean and shade for the $25^{\rm th}$--$75^{\rm th}$ 
    percentiles among all galaxies in our sample). For comparison,
    we also show the same parameters for all stars (blue).
    The time variable of each galaxy is measured with respect to
    $t^{(\rm start)}_{\gamma}$ and normalized by
    $\Delta t_{\gamma} \equiv t^{(\rm end)}_{\gamma} - t^{(\rm start)}_{\gamma}$.
    Grey vertical band in each panel indicates the transitional regime of halo assembly.
    {\figem (e)}, the migration of bulges in the $c/a$-$b/a$ plane,
    with each pair of empty and filled markers linked by a line 
    representing the bulge of a galaxy in the hot and cold phases, 
    respectively. The evolution of bulge structure implies 
    the two distinct modes of star formation that drive the dynamical
    hotness, which may be tested by JWST and future observations.
    See \S\ref{sec:early-growth} for details.
    }
    \label{fig:bulge_evolution}
\end{figure*}

\begin{figure*}
	\includegraphics[width=0.8\textwidth]{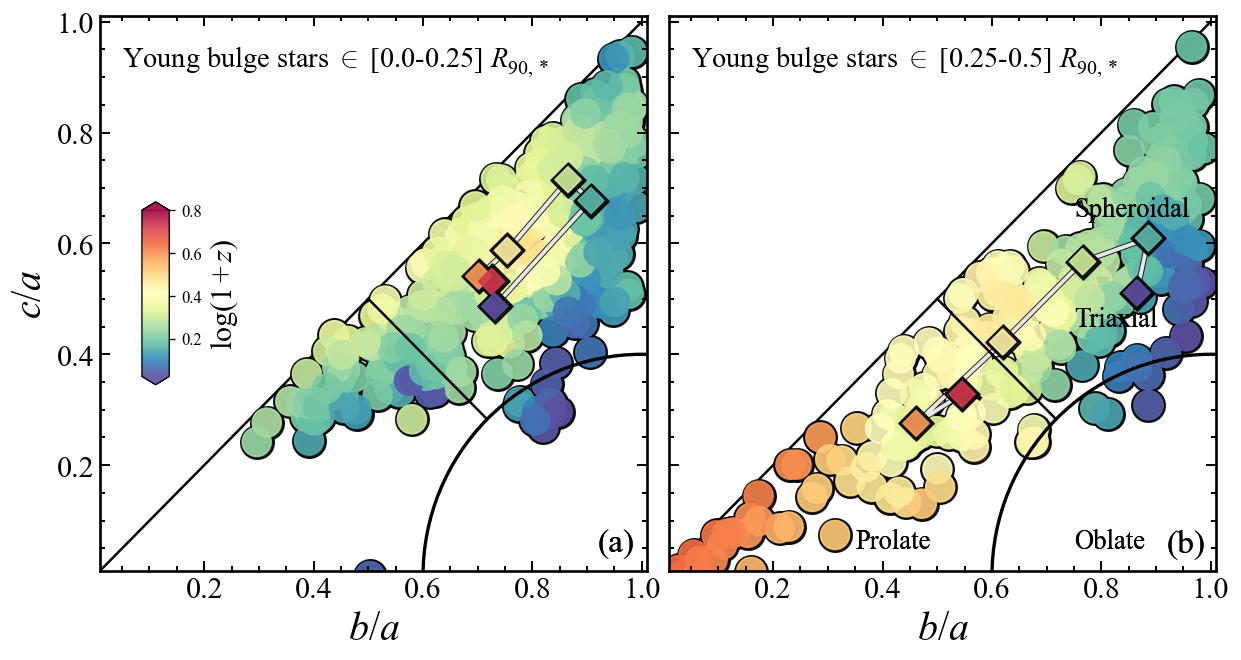}
    \caption{{\figem Distribution of stellar bulges in the $b/a$-$c/a$ plane.}
        Here we only include young bulge stars in given ranges of 
        galactocentric distance: {\figem (a)}, $r < 0.25 R_{\rm 90,*}$;
        {\figem (b)}, $0.25 R_{\rm 90,*} < r < 0.5 R_{\rm 90,*}$.
        Data points show the snapshots of all galaxies in our sample,
        with color coded according to the redshift smoothed by the LOESS method \citep{Cleveland01091988}.
        In each panel, squares show the evolution of the axis ratios 
        obtained by the median of all galaxies.
        This figure verifies that scattered star formation in high-$z$ 
        galaxies takes place at large radii in cold streams, thus producing 
        anisotropic shapes of bulges. See \S\ref{sec:early-growth} 
        for details.
    }
    \label{fig:bulge_shape_origin}
\end{figure*}

\section{Structural evolution of dynamically hot stellar systems}
\label{sec:early-growth}

A particularly interesting part of the two-phase formation of galaxies is the 
growth of dynamically hot stellar systems (referred to as bulges in this paper)
in fast-accreting halos. The importance of bulge formation is threefold. First, the 
`scattered formation' of stars in the early epoch depends on the competition between 
various cooling and heating processes. Many of these processes are unresolved in current
simulations and are implemented as subgrid models. The bulge components 
formed through this channel may thus encode information about these
processes, and their properties may be used to distinguish models.
Second, as shown in Fig.~\ref{fig:quadrant}(e) and other 
previous studies using different simulations \citep{pillepichFirstResultsTNG502019,yuBornThisWay2023}, 
the bulge components dominate the total stellar mass of high-$z$ galaxies, 
and thus their properties have important implications for observations targeting 
high-$z$ galaxies. Recent observations with resolved stellar dynamics reveal surprising cases 
of disk-like galaxies at $z \gtrsim 4$, which appears at odd with theoretical expectations.
Third, bulges are found to co-evolve with the central supermassive black holes (SMBHs) of galaxies
\citep{kormendyCoevolutionNotSupermassive2013,greeneIntermediateMassBlackHoles2020,
zhuangEvolutionaryPathsActive2023}, and thus closely related to understanding 
galaxy quenching \citep{dekelWetDiscContraction2014,wooTwoConditionsGalaxy2015,linSDSSIVMaNGAInsideout2019,
hongDynamicalHotnessStar2023,moTwophaseModelGalaxy2024}. The bulges in FIRE-2 
may thus be used to study potential sites of SMBH growth.
Here we focus on bulges in FIRE-2 by identifying properties that not only 
reflect the physical processes driving the formation and evolution, but 
also can be measured observationally.

Within the two-phase model,  \citet{chenTwophaseModelGalaxy2025} showed that 
the fraction of star formation in dense star clusters is expected to increase 
towards high $z$ \citep[][see their fig. 18]{chenTwophaseModelGalaxy2025},
and such a trend can be used to explain the formation of old globular clusters.
To see if this expectation is supported by FIRE-2 galaxies, we show,
in Fig.~\ref{fig:bulge_illustration}(a)--(c), the projected stellar distribution of 
the bulge component of the galaxy {\objname m12f} at three redshifts chosen to 
represent different stages of bulge formation. The change in the clumpiness 
with time is obvious: the galaxy is highly clumpy at $z = 4.4$, becomes less clumpy 
at $z = 1.5$, and appears smooth at $z = 0$. A similar finding was also 
reported by \citet{maSelfconsistentProtoglobularCluster2020} 
for a number of other galaxies in the FIRE-2 suite.

Aided by strong lensing, recent observations by JWST revealed an excess of high-$z$ galaxies 
that are dominated by dense stellar clumps
\citep[e.g.][]{vanzellaJWSTNIRCamProbes2023,linMetalenrichedNeutralGas2023,mowlaFireflySparkleEarliest2024,
adamoBoundStarClusters2024,Fujimotoz9disk2025}, supporting the expectation that
stellar dynamics is hot in high-$z$ galaxies. Unfortunately, most
galaxies without being strongly lensed are not well resolved, and so it is difficult 
to make a conclusion for the overall galaxy population.
Even more challenging is to quantify potential selection bias in such observations. 
Similar limitations in resolution and selection exist for observations that depend on resolved gas 
dynamics \citep[e.g.][]{parlantiALMAHintsPresence2023,
rowlandREBELS25DiscoveryDynamically2024}, as the selection is naturally biased toward galaxies 
with large masses and significant rotation.

Given the different formation channels of dynamical hotness found in 
\S\ref{ssec:origin-hotness}, we expect that the structural properties of bulges
should evolve with time. Bulge sizes, which have been discussed in \S\ref{ssec:origin-hotness} 
and Fig.~\ref{fig:radius}, are expected to be large during the scattered 
formation at the fast accretion phase, and to stop growing or even shrink 
during the later concentrated formation. Bulges at high $z$ are also expected to show 
significant elongation in shape and small concentration in radial profile.
The expectation of elongated shapes for high-$z$ galaxies is based on the 
scattered star formation in dense gas clumps associated with cold streams
\citep[e.g.][]{mandelkerColdFilamentaryAccretion2018,danovichFourPhasesAngularmomentum2015}. Stars formed in such a
way are expected to inherit the overall velocity field of dominating 
streams, making the stellar distribution preferentially along the major
axis of the velocity ellipsoid. The low concentration is expected directly
from the `scattered' nature of star formation together with subsequent dynamical
heating that makes the stellar distribution more extended. 
In contrast, bulge formation at lower $z$ is mainly fed by the hot-mode 
accretion that is more isotropic, with larger amounts of dissipation 
of binding energy in the halo.  Dynamical heating is also less significant at low
$z$, all together making the bulges more spheroidal and more concentrated. 

The evolution of the bulge structure can be seen from the example images shown in 
Fig.~\ref{fig:bulge_illustration}(a)--(c).
At $z=4.4$, the stellar distribution is elongated along the diagonal direction
of the image, exhibiting filament-like structures that trace the accretion flow 
of gas into the galaxy. We visually inspected the cold gas stream and found that 
it is indeed aligned with the stellar distribution. Meanwhile, the stellar distribution 
is extended. 
The size of the bulge (indicated by a dashed circle; see also Fig.~\ref{fig:radius}a) is already comparable to the size at $z=0$, 
although the stellar mass is much lower than the final bulge mass (see Fig.~\ref{fig:Example}a). 
The concentration of stars in the galaxy is evident, but the multiple dense 
stellar clumps near the center make it difficult to identify the center.
At $z=1.5$, the stellar distribution is less elongated, with
a well-defined concentration. At $z = 0$, the bulge looks spherical, and the 
concentrated core is more prominent. In panels (d)--(f), we show the evolution of 
the stellar mass surface density profile [$\Sigma_*(r_{\rm p})$],
the S\'ersic index ($n$), and the shape parameters
(axis ratio, $c/a$, and the triaxiality, $T$), respectively,
of this bulge.
As can be seen, the bulge evolves from a prolate, low-concentration structure at high $z$ 
to a spheroidal, high-concentration structure at low $z$.
There appears to be a clear transition in the evolution of the structural parameters, 
and the transition time corresponds well to that in halo assembly
($t_\gamma$, marked by the red solid lines in panels e and f) and in 
dynamical hotness of the galaxy ($t_{\cal H}$, 
marked by the red dashed lines in panels e and f).
This provides further support that the evolution of bulges is consistent with being influenced by
the evolution of dark matter halos.

In Fig.~\ref{fig:bulge_evolution}, we show the evolution of the bulge structure
for all the 14 galaxies in our sample.
In panels (a)--(d), we show the evolution of the structural parameters: 
S\'ersic index $n$, triaxiality $T$, $b/a$ and $c/a$, respectively,
as functions of cosmic time 
for the median and $25^{\rm th}$--$75^{\rm th}$ percentiles among all bulges. 
All these results confirm the expectation that the bulge structure
evolves from being prolate (small $c/a$ and $b/a$; large $T$) to being 
spheroidal (large $c/a$ and $b/a$; small $T$), and the bulge concentration (S\'ersic index $n$) 
increases with time. The scatter of $n$ is large at high $z$,
reflecting the presence of violent dynamical disturbances that 
can distort the distribution of stars and prevent the formation of a 
well-defined galactic center (see also Fig.~\ref{fig:bulge_illustration}c and f). 
We checked the S\'ersic fitting and found that 
the $\chi^2$ is significantly enlarged at $z\gtrsim 2$, 
again indicating the distorted stellar distribution and the 
difficulty of using a single analytical profile to capture such distortion. 
The cosmic times in panels (a)--(d) are aligned with $t_{\gamma}^{\rm (start)}$ 
for individual galaxies. Clearly, the transition in shape properties is closely related 
to the transition of the halo assembly. This tight relation between the transition in bulge 
shape and the transition in halo assembly seems to indicate that the accretion rate 
and accretion configuration are correlated with each other, so that the change
from fast to slow accretion of the halo corresponds to the change from 
an anisotropic cold-mode accretion to an isotropic hot-mode accretion of gas into the 
galaxy. In panel (e), we show the locations of bulges in the $c/a$--$b/a$ plane 
at two epochs: one is the earliest time when a reliable measurement of the shape 
parameters can be made for a galaxy (when the number of the star particles is larger than 
1000) in the dynamically hot phase, and the other is at $z = 0$, when the halo is already 
in slow accretion. We see clearly that the evolution is from being prolate 
to being spheroidal \citep[see also][]{mengEvolutionDiscThickness2021,XiaFire3DShape2025}. 

To verify the expectation that the prolate shape of bulges at high $z$ is
indeed an outcome of the early scattered star formation in anisotropic,
cold gas streams, we divide newly formed bulge stars, defined to be
ones with ages in the lowest quarter of the age distribution
(to ensure a reasonably large sample size), into subsamples according to the galactocentric 
distance, and measure the shape parameters using stars in each subsample.
If early scattered formation is the main driver of the prolate shape,
bulge stars formed at large radii should be more prolate than those formed at small 
radii where streams are mixed with pre-existing gas to form a more isotropic distribution.
On the other hand, if other factors, such as dynamical heating, are the main driver,
young stars should not show significant anisotropy in their distribution.
In Fig.~\ref{fig:bulge_shape_origin}, we show galaxies in the  
$c/a$-$b/a$ diagram for young bulge stars formed in two different radial ranges:
$0$--$0.25 R_{90,*}$ and $0.25R_{90,*}$--$0.5R_{90,*}$. 
Clearly, young bulge stars 
at large radii show more prolate shape than those formed at smaller radii,
confirming that the prolate shape of bulges at high $z$ is an outcome of the early 
scattered star formation in an anisotropic structure.

In observations, resolving structural properties such as the shape and concentration 
is less demanding than resolving small-scale clumpiness. The evolution of the bulge structure thus 
provides a promising way to test the two-phase formation model by observations.
Indeed, recent observations of high-$z$ galaxies with JWST indicate an excessively large fraction of 
galaxies with prolate stellar distribution \citep{LawHSThighzstellardisk2012,YumaGoodzhighzbarlike2012,VanderWelGeometryStarforming2014,ZhangCANDELSprolate2019,PandyaJWST3Dprolateness2024}. However, the small S\'ersic index shown 
in Fig.~\ref{fig:bulge_evolution}(a) for FIRE-2 galaxies at high $z$ indicates that the measurements 
of the S\'ersic index may not be able to distinguish prolate ellipsoids from inclined disks.
Caution is thus needed when interpreting the density profiles of high-$z$ galaxies.  
For comparison, we show the evolution of the median of individual structural parameters 
using all stars in Fig.~\ref{fig:bulge_evolution}(a)--(d). Except for the S\'ersic index, 
the evolution of the structural parameters is less significant than that obtained 
using only bulge stars, suggesting that a precise decomposition between the bulge 
and disk components is needed to constrain the formation mechanism of dynamically hot stellar systems.

\section{Summary and discussion} 
\label{sec:summaryanddiscussion}


In this paper, we use 14 MW-size galaxies from the FIRE-2 cosmological 
zoom-in simulations to study the co-evolution between galaxies and their 
host dark matter halos. We focus on the transition behavior of galactic 
dynamics that separates the evolution into two distinct phases, 
and we link the two-phase formation of galaxies to 
that of their host halos.
We particularly identify the processes from which dynamical hotness 
of galaxies originates, and we study the implications of our results
for testing these processes using observations. Our main results and conclusions 
are summarized as follows.
\begin{enumerate}[topsep=0pt,parsep=0pt,itemsep=0pt]
    \item We define a set of parameters that either quantify the dynamical 
    state of a galaxy or reflect the consequence of a given dynamical state. These include
    the stellar masses in different kinematic components 
    (thin disk, thick disk and bulge), the mass fractions of these components 
    in the total stellar mass of a galaxy
    ($f_{\rm thin}$, $f_{\rm thick}$ and $f_{\rm bulge}$),
    the dynamical hotness ($\mathcal{H}$), and
    the burstiness of star formation ($\mathcal{B}$).
    We find that all these parameters follow a two-phase pattern in their 
    time evolution, 
    and we identify the characteristic times for the evolution of these parameters:
    the half-mass formation time of the bulge ($t_{\rm bulge,1/2}$),
    the transition time when mass fraction of the thin-disk component 
    increases rapidly with time ($t_{\rm thin}$), the transition time 
    in dynamical hotness ($t_{\cal H}$) of a galaxy, and the transition time  
    in the burstiness of star formation ($t_{\cal B}$)
    (\S\ref{ssec:galaxy-properties}; Figs.~\ref{fig:visualization} and \ref{fig:Example}).
    
    \item
    We investigate the co-evolution between galaxies and their 
    halos by linking each of the above characteristic times of galaxies to
    the time at which the assembly of halo transits from the fast phase to the slow
    phase ($t_\gamma$), and we find that all the characteristic times of galaxies
    correlate strongly with $t_\gamma$. This motivates us to explore the 
    causal connection between the two-phase formation of halos and 
    that of galaxies.
    The transition of $f_{\rm thin}$ and $\mathcal{H}$ usually completes 
    within $1$--$3$ dynamical timescales of the host halo, 
    with a few outliers 
    that take longer time to complete the dynamical adjustment. 
    (\S\ref{sec:co-evolution}; Fig.~\ref{fig:Correlation}).
    
    \item
    Motivated by \citetalias{moTwophaseModelGalaxy2024}, we use 
    two quantities, the specific growth rate of halo ($\gamma$)
    and the ratio between cold-gas fraction and cold-gas spin 
    ($f_\lambda$)
    to link the evolution of galactic dynamics to the underlying mechanisms.
    The critical values, $\gamma = \gamma_{\rm f} \approx 0$--$3/8$
    and $f_\lambda = 1$, separate the evolution of galaxies 
    into four quadrants (Q1--Q4) in the $f_\lambda$-$\gamma$ plane.
    Galaxies in each quadrant are expected to have distinct 
    dynamical features, and the transitions at around these critical values
    of the FIRE-2 galaxies are consistent with the expectation. 
    The general evolution trend of galaxies in the quadrant diagram is 
    from Q1 (high $\gamma$ and high $f_\lambda$) to Q3 
    (low $\gamma$ and low $f_\lambda$), with the initial state 
    set by the $\Lambda$CDM cosmology and the evolution pattern 
    shaped by the synergy of the stall 
    of rapid halo assembly and the reduction of gas fraction by feedback (\S\ref{ssec:quadrant};
    Fig.~\ref{fig:quadrant}).

    \item 
    The dynamical hotness of galaxies originates from two distinct modes 
    of star formation. The first one is a scattered mode, occurring at 
    $\gamma \gtrsim \gamma_{\rm f}$, in which stars form out of 
    locally unstable gas clumps at large radii, coinciding with the existence of gaseous streams 
    seeded by the dark matter filaments of the cosmic web.
    The second one is a concentrated mode,
    occurring at $\gamma \approx \gamma_{\rm f}$,
    in which stars form out of sub-clouds fragmented from the entire 
    self-gravitating gas cloud (SGC).
    The large amount and high density of the accumulated gas, and 
    the deepened gravitational potential well, 
    make the star formation in the concentrated mode highly violent and lead 
    to the formation of the majority ($\sim 70\%$) of bulge stars  
    seen at $z = 0$. 
    The migration of star formation 
    from the scattered mode to the concentrated mode
    is inferred to be caused by the disappearance of
    gaseous streams and the cease of dynamical heating
    in slow-accreting halos (\S\ref{ssec:origin-hotness};
    Figs.~\ref{fig:Bulge_mass_tranistion} and \ref{fig:radius}).
    
    \item
    At $\gamma < \gamma_{\rm f}$, the perturbation of gravitational  potential 
    by halo accretion ceases. 
    Concentrated star formation has significantly reduced the fraction 
    of gas via violent feedback in the inner SGC 
    (Fig.~\ref{fig:Bulge_mass_tranistion}), 
    and built up a concentrated gravitational
    potential dominated by the bulge stars,
    both preventing the gas from self-gravitating globally.
    Dynamically cold gaseous and stellar disk now have enough time to form.
    The growth of the bulge via the concentrated mode continues, but 
    is limited to the innermost region where self-gravitation of gas 
    can still be achieved  (\S\ref{ssec:process-transition}).

    \item 
    The two modes of star formation may leave observable 
    imprints on the structures of bulges. 
    Bulge stars formed in the scattered mode have large spatial 
    extents (a few tens percent of $R_{\rm vir}$) and small concentration,
    and are distributed in a prolate shape elongated along the
    inflow stream.
    In contrast, bulge stars formed in the concentrated mode
    have small spatial extents 
    (a few percent of $R_{\rm vir}$) and large concentration,
    and are distributed in a spheroidal shape. Other observable 
    signatures related to galactic dynamics, such as the
    burstiness of star formation, and the clumpiness of star-forming gas
    and young stars, may also be used to decode the mechanisms
    driving the dynamical evolution of galaxies
    (\S\ref{sec:early-growth}; Figs.~\ref{fig:bulge_illustration}, \ref{fig:bulge_evolution} and \ref{fig:bulge_shape_origin}).
\end{enumerate}

The findings here support the two-phase formation scenario of galaxies and
the use of halo assembly to discriminate the two phases of galaxies.
However, FIRE-2 is specialized by its subgrid physics model, 
leading to lack of the compact, baryonic-dominated bulges of the high 
redshift dwarfs \citep{ShenDissipativeDMFIRE2024} and too dispersion-dominated 
galaxies for the at intermediate masses ($M_{\rm *} \sim 10^{\text{8--10}} \rm M_{\odot}$;
see e.g. \citealt{ElBadryFIREIIgaskinemati2018,KadoFongFIREdwarfstellaroutskirts2022}). 
Some of our conclusions may depend on the sub-grid models adopted in FIRE-2, 
For example, as noted at the end of \S\ref{ssec:origin-hotness}, the
`early' feedback produced by young stars and the external feedback from 
the UV background may be able to disperse the gas in cold filaments,
thereby reducing the amount of stars that can form in the scattered mode
\citep[spurious heating term at $z\gtrsim 8$, see e.g.][]{SuFIREdwarfsne2018,GarrisonDwarfsatellitessfr2019}. 
Numerical resolution also matters. The formation of dense gaseous
and stellar clumps from the turbulent medium may continuously interact with the surrounding gas
and stars, causing significant dynamical heating that limits star
formation in concentrated mode \citep[e.g.][]{moGalaxyFormationPreprocessed2004,
ogiyaDarkMatterCores2022,
mccluskeyDiscSettlingDynamical2024}. Failing to
properly resolve such clumps may therefore lead to incorrect results.  
Another issue arises from the specific sample of galaxies selected by the
zoom-in runs of FIRE-2, which imposes isolation criteria on the target halos
that can bias the sample against dense environments. 
In Appendix~\ref{app:sample_selection}, we show the distribution of galaxies
in the $t_{\rm h,1/2}$-local density plane, and we find that the
bias in $t_{\rm h,1/2}$ is small, but that in local density may
be significant. Clearly, all these issues need to be studied further
using simulations with improved sub-grid models and samples with high completeness.

Another important component that is absent in our analysis is 
supermassive black holes (SMBHs). As suggested by the analytical estimation in
\citet[see their \S4.3 and \S4.4]{chenTwophaseModelGalaxy2025a}, 
the turbulence of gas plays a fundamental role in breeding 
the seeds of SMBHs, promoting their early growth, and determining the
nuclear structure of galaxies. The growth of SMBHs is thus expected to be
closely related to the build-up of the bulge components of galaxies in their hot phase.
The rapid growth of bulge during the transitional phase
through the concentrated mode coincides with the `blue nugget' phase
found in observations \citep{BarrCompactSFGsCANDELS2014,
BarroFormationDenseGalactic2016,
jiReconstructingAssemblyMassive2023}, which was inferred to be 
closely related to the boost of AGN activity 
\citep{carrIdentificationQuenchingNugget2024,cohnEvidenceEvolutionaryPathwaydependent2025,
carrUsingMachineLearning2025,
chenTwophaseModelGalaxy2025a}.
The co-evolution among SMBHs, galaxies, and halos should thus be
investigated coherently to form a complete picture of galaxy formation.
We will return to such an investigation using hydro simulations
that include all these components.

While our correlation analyses reveal a co-evolution pattern 
between galaxies and halos, a causal direction between the two has not yet
been firmly established. In fact, systematic changes in other properties of galaxies, 
such as a drop in the burstiness of star formation 
(\citealt{muratovGustyGaseousFlows2015,AnglesAlcazarCosmicBaryon2017,
wangVariabilityStarFormation2020,wangVariabilityStarFormation2020a,
sternVirializationInnerCGM2021,yuBurstyOriginMilky2021,yuBornThisWay2023}; 
see also Fig.~\ref{fig:Correlation}), the steepening of the shape of potential well
(\citealt{hopkinsWhatCausesFormation2023,semenovHowDiskGalaxies2026}; 
see also our discussion in \S\ref{ssec:process-transition}), 
and the thermalization of the CGM (\citealt{sternVirializationInnerCGM2021,
hafenHotmodeAccretionPhysics2022,GurvichFIREdiskSFR2022,KakolyTurbulencedominatedCGMOrigin2025}; 
see also our discussion in \S\ref{ssec:process-transition}),  all occur around an epoch similar to the transition in 
dynamical hotness of galaxies. Disentangling these processes, identifying their synergies
and determining the causal sequence require controlled simulations, in which a subset of processes 
are activated at each time, and detailed analysis of a large ensemble of simulations that encompasses 
different combinations of relevant processes.

Our results have important implications for observations to 
search for the evidence of scattered star formation using structural properties 
(size, axis-ratios, triaxiality and S\'ersic index) of bulges.
Current observations with JWST are already capable of measuring
some of these properties for high-$z$ galaxies
\citep{LawHSThighzstellardisk2012,YumaGoodzhighzbarlike2012,VanderWelGeometryStarforming2014,
ZhangCANDELSprolate2019,PandyaJWST3Dprolateness2024,jiaSizeGrowthShort2024,songTransitionOutsideinInsideOut2025}. 
However, because of the flux limit, the measurements may miss the faint outskirts
of galaxies that are expected to carry the most significant
imprints of anisotropic star formation (Fig.~\ref{fig:bulge_shape_origin}).
On the other hand, identifying the concentrated mode of star formation
observationally may rely on the compact galaxy morphology and rapid rise in SFR
associated with this mode. At low $z$,
\citet{wangElevationSuppressionStar2019} used the MaNGA dataset and found that
compact galaxies are more broadly distributed around the star-formation main sequence
than extended galaxies at fixed $M_*$ \citep[see also][]{wuytsGalaxyStructureMode2011,
wangElevationSuppressionStar2019,heSymmetryFundamentalParameters2026}.
This behavior has been attributed to differences in the temporal variability of
their SFRs, an interpretation further supported by measurements of galaxy burstiness
using SFRs over different timescales \citep{wangVariabilityStarFormation2020,wangVariabilityStarFormation2020a}.
However, these results are currently limited to low-$z$ galaxies.
A stacked analysis of a large sample of galaxies may thus be necessary to
reveal the full extents of high-$z$ galaxies and to put constraints on the
formation mechanisms of galaxies in the early Universe.

\section*{Acknowledgements}

QM thanks Cheng Li, Tao Jing, Yunwei Deng and Yunjing Wu for discussions.
YC is funded by 
the National Natural Science Foundation of China (Grant no. 12503014),
the Fundamental Research Funds for the Central Universities (Grant no. KG202502)
and the China Postdoctoral Science Foundation (Grant no. 2022TQ0329),
and thanks Hui Li and Kai Wang for their valuable insights
and suggestions. 
The authors also acknowledge the Tsinghua Astrophysics High-Performance 
Computing platform at Tsinghua University for computational resources, 
and Yi Mao and Chen Chen for technical support.

\section*{Data Availability}

Data and codes used to produce the results in this paper are available 
upon reasonable request to the corresponding authors.
Public data of the FIRE project is available at
\url{http://flathub.flatironinstitute.org/fire}.

\bibliographystyle{mnras}
\bibliography{ref.bib}

\appendix

\section{The sample of FIRE-2 galaxies}
\label{app:sample_selection}

\begin{figure*}
	\includegraphics[width=0.96\textwidth]{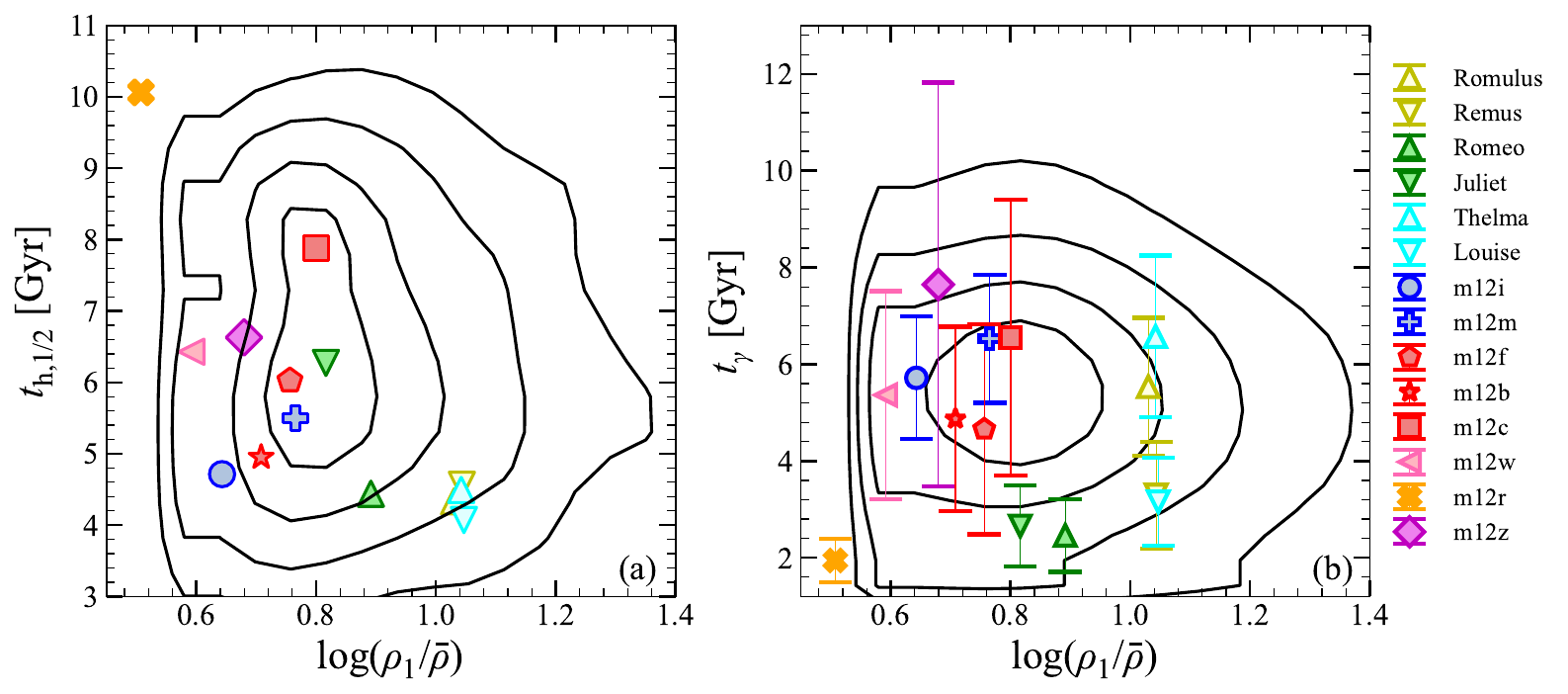}
    \caption{{\figem Distribution of FIRE-2 galaxies in the assembly-environment planes.} 
    Each colored marker shows the location of a FIRE-2 galaxy at $z=0$
    in {\figem (a)}, the $t_{\rm h,1/2}$--$\rho_1/\bar{\rho}$ plane; 
    {\figem (b)}, the $t_{\gamma}$--$\rho_1/\bar{\rho}$ plane.
    Here, $\rho_1/\bar{\rho}$ is the local density contrast within
    $1\mpc$ around each galaxy. In each panel,
    black contours, from inner to outer, encompass 25\%, 
    50\%, 75\% and 90\% of halos with masses comparable to the FIRE-2 hosts,
    taken from TNG300-1-Dark, a dark-matter-only simulation 
    covering a large volume.
    The FIRE-2 galaxies are not significantly biased in terms of
    halo assembly at given local density, but those in 
    dense environments are missed.
    See Appendix~\ref{app:sample_selection} for details.
    }
    \label{fig:Compare_largesim}
\end{figure*}

As briefly introduced in \S\ref{sec:sim}, the sample of FIRE-2 galaxies used in 
this study is drawn from a cosmological simulation, with selection criteria 
applied to ensure the isolation or the mimicking of the Local Group environment.
A question thus arises as to whether our conclusions on the co-evolution 
between galaxies and halos are biased by the sample selection.
Here we address this question by comparing the distribution of our sample
with a complete sample drawn from a large-volume cosmological simulation.

For this purpose, we use TNG300-1-Dark, a dark-matter-only simulation
performed as a part of the IllustrisTNG project
\citep{pillepichFirstResultsIllustrisTNG2018, 
nelsonIllustrisTNGSimulationsPublic2019}. 
This simulation has a box size of about $300\Mpc$ on a side, 
and a mass resolution that can resolve halos with 
$M_{\rm vir} \gtrsim 2 \times 10^9\Msun$. From this simulation, we select
a `reference sample' that includes all
halos at $z=0$ with $M_{\rm vir}$ in the range of $10^{12}$--$10^{12.05}\msun$, 
comparable to the MW-size halos hosting the FIRE-2 galaxies.
For halos in FIRE-2 and the reference sample, we calculate $\rho_1$, 
the local matter density within 
a sphere with a radius of $1\,h^{-1} {\rm Mpc}$ centered on each halo,
and normalize it by the mean density of the universe at $z=0$ 
to get the local density contrast, $\rho_1/\bar{\rho}$.
We also calculate the half-mass formation time, $t_{\rm h,1/2}$,
and transition time, $t_{\gamma}$, for each halo, 
following the definitions in \S\ref{ssec:halos}. We show the distribution of FIRE-2 halos and 
TNG300-1-Dark halos in the $t_{\rm h,1/2}$-$\rho_1/\bar{\rho}$
and $t_{\gamma}$-$\rho_1/\bar{\rho}$ planes 
in Fig.~\ref{fig:Compare_largesim}.

All the FIRE-2 halos are found to be embedded in environments with 
local density contrast $\rho_1/\bar{\rho} \lesssim 10$.
Paired systems have higher local densities than the isolated ones.
The reference sample extends to higher densities of 
$\rho_1/\bar{\rho} \gtrsim 10^{1.3}$.
This suggests that the selection in FIRE-2 is indeed biased towards 
under-dense environments, as is expected given the isolation criteria used 
for the isolated galaxies in the simulation.
However, at a given local density, the distributions of $t_{\rm h,1/2}$ 
and $t_{\gamma}$ do not show significant differences between
the FIRE-2 sample and the reference sample, implying that
the selection does not significantly bias the sample in terms of 
halo assembly. The analyses in this paper thus fairly
represent halos that are sufficiently isolated, while halos
susceptible to strong environmental effects due to the proximity
to massive cosmic structures are missed.

\section{The state of inner circumgalactic medium}
\label{app:CGM-virialization}

Following \citet{sternVirializationInnerCGM2021}, we use two timescales to
quantify the state of the CGM: the cooling time of shocked gas,
$t_{\rm cool}^{\rm (s)}$, and the local free-fall time, $t_{\rm ff}$.
A ratio of $t_{\rm cool}^{\rm (s)}/t_{\rm ff} \lesssim 1$ indicates that
shocked gas cools rapidly, loses pressure support, and free-falls toward the
galaxy, whereas $t_{\rm cool}^{\rm (s)}/t_{\rm ff} \gtrsim 1$ indicates that
shocked gas can remain virialized at approximately the virial
temperature of the halo.

The local free-fall time at a given galactocentric distance $r$ is
\begin{equation}
    t_{\rm ff} = \sqrt{2}r/v_{\rm c}
    \,,
\end{equation}
where $v_{\rm c}(r) \equiv \sqrt{GM(<r)/r}$ is the circular velocity at $r$,
and $M(<r)$ is the total mass enclosed within $r$.
The cooling time is defined as the time that gas would have if it were shock-heated
to the virial temperature and remained pressure-supported:
\begin{equation}
    t_{\rm cool}^{\rm (s)} 
    \equiv \frac{(3/2) \times 2.3 k_{\rm B} T^{\rm (s)}}{n_{\rm H}^{\rm (s)} 
    \Lambda \left(T^{\rm (s)}, n_{\rm H}^{\rm (s)}, Z\right)}
    \,.
\end{equation}
Here, $k_{\rm B}$ is the Boltzmann constant, $Z$ is the gas metallicity in a
spherical shell around $r$, and $\Lambda$ is the cooling function obtained by
interpolating the tables of \citet{wiersmaEffectPhotoionizationCooling2009}. 
The temperature $T^{\rm (s)}$ is derived as
\begin{equation}
    T^{({\rm s})} \equiv \frac{3 \mu m_{{\rm p}} v_{{\rm c}}^2}{5 A k_{{\rm B}}}
    = 4.5 \times 10^5 A^{-1} \left(\frac{v_{\rm c}}{100\kms}\right)^2 \Kelvin
    \,,
\end{equation}
where $\mu = 0.62$ is the mean molecular weight, $m_{\rm p}$ is the proton
mass, and
\begin{equation}
    A=\frac{9}{10}\left(1-2 \frac{d \log v_{{\rm c}}}{d \log r}\right) \approx 1
    \,.
\end{equation}
$n_{\rm H}^{\rm (s)}$ is the hydrogen number density, obtained by balancing
the gas pressure against the gravity of overlying material:
\begin{equation}
    2.3 n_{\rm H}^{\rm (s)} k_{\rm B} T^{\rm (s)} = \int_{r}^{R_{\rm vir}} 
    \frac{\rho(r') v_{\rm c}^2(r')}{r'} dr'
    \,,
\end{equation}
where $\rho(r')$ is the spherically-averaged gas density at $r'$ obtained
from the simulation. We measure $t_{\rm cool}^{\rm (s)}$ and $t_{\rm ff}$
in the inner CGM at $r = 0.1 R_{\rm vir}$ since this region is directly
relevant to gas accretion onto the galaxy.

\begin{figure}
	\includegraphics[width=0.48\textwidth]{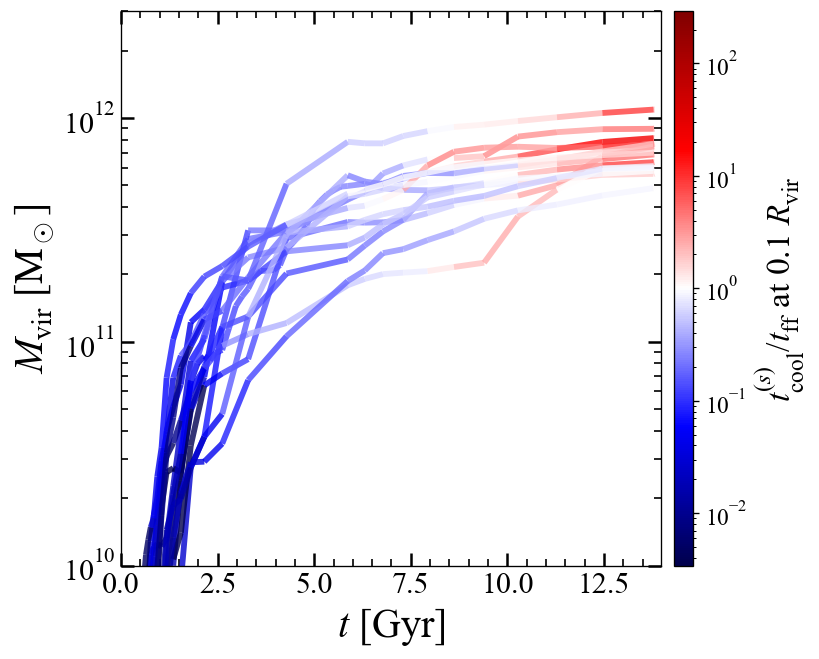}
    \caption{{\figem CGM virialization in FIRE-2 galaxies.} Here we 
    show halo mass ($M_{\rm vir}$) as a function of cosmic time ($t$) for each of the 14 FIRE-2 galaxies, 
    colored by the ratio of the cooling time of shocked 
    gas to the local free-fall time at $0.1 R_{\rm vir}$. The inner CGM of all 
    galaxies virializes in the slow-accretion phase, as indicated by 
    $t_{\rm cool}^{\rm (s)} / t_{\rm ff} \gtrsim 1$.
    See Appendix~\ref{app:CGM-virialization} for details.
    }
    \label{fig:tcool_tff_Mhalo}
\end{figure}

Fig.~\ref{fig:tcool_tff_Mhalo} shows the evolution of $M_{\rm vir}$ for
the individual galaxies in our sample, colored by the ratio
$t_{\rm cool}^{\rm (s)}/t_{\rm ff}$ at $0.1 R_{\rm vir}$.
We find that, for all galaxies, the inner CGM is not virialized when their
host halos are in the fast-accretion phase. One condition necessary for the
build-up of dynamical hotness in this phase, namely efficient cooling of the
shocked gas, is therefore satisfied during fast accretion. However, the inner
CGM appears close to virialization, with $t_{\rm cool}^{\rm (s)}/t_{\rm ff}
\sim 1$, in the transitional phase (white segments in the figure).
Inner-CGM virialization may thus provide an alternative or complementary 
mechanism to control the transition in dynamical hotness of galaxies.

\bsp	
\label{lastpage}
\end{document}